\pdfoutput=1
\documentclass[a4paper,times,3p]{elsarticle}
\usepackage{lineno}

\usepackage{amsmath}
\usepackage{xcolor} 	
\usepackage[]{algorithm2e}
\usepackage{tabularx}
\usepackage{multirow}
\usepackage{subfig}

\newcommand{\vekt}[1]{\mathbf{#1}}
\newcommand{\bolditalic}[1]{\textit{\textbf{#1}}} 
\newcommand{\textitbf}[1]{\textit{\textbf{#1}}} 
\newcommand{\textbfit}[1]{\textit{\textbf{#1}}} 
\newcommand{\diff}[1]{\nu \vekt{\nabla}^2 \vekt{v} }
 
\newcommand{\vel}[0]{\vekt{u}^f}

\newcommand{\Rtil}[0]{\vekt{\tilde{R}}}
\newcommand{\xtil}[0]{\vekt{\tilde{x}}}
\newcommand{\res}[1] {\vekt{R}(\vekt{x}_d^{#1})}
\newcommand{\Jac}[0] { \widehat{\vekt{J}}_{-1}}
\newcommand{\JacN}[1] { \mbox{}^{#1} \widehat{\vekt{J}}_{-1}}
\newcommand{\CauchyStress}[0]{\vekt{T}}
\newcommand{\FluidStressK}[0]{\CauchyStress_k^f}
\newcommand{\nsd}[0]{{nsd}}
\newcommand{\fluidSolver}[0]{{\boldsymbol{\mathcal{F}}}}
\newcommand{\structSolver}[0]{{\boldsymbol{\mathcal{S}}}}
\newcommand{\fixPntOperator}[0]{{\boldsymbol{\mathcal{H}}}}
\newcommand{\Wk}[0] { \vekt{W}_k}
\newcommand{\Vk}[0] { \vekt{V}_k}
\newcommand{\Zk}[0] { \vekt{Z}_k}

\newcommand{\Vhrulefill}[0] {\leavevmode\leaders\hrule height 0.7ex depth \dimexpr0.4pt-0.7ex\hfill\kern0pt}
\newcommand{\comment}[1] { {#1} }
\newcounter{para}
\newcommand \mypara{\refstepcounter{para} \par \noindent\textit{Remark \thepara:\space}}

\journal{Computer Methods in Applied Mechanics and Engineering}

\begin{document}

\begin{frontmatter}



\title{A Multi-Vector Interface Quasi-Newton Method with Linear Complexity for Partitioned Fluid-Structure Interaction}

 \author{Thomas Spenke\corref{cor1}}
 \ead{spenke@cats.rwth-aachen.de}
 
 \author{Norbert Hosters}
 \ead{hosters@cats.rwth-aachen.de}
 
  \author{Marek Behr}
  \ead{behr@cats.rwth-aachen.de}
 
 \cortext[cor1]{Corresponding author}
 
 \address{Chair for Computational Analysis of Technical Systems (CATS), Center for Computational Engineering Science (CCES),\\RWTH Aachen University, 52062 Aachen, Germany}

\begin{abstract}

In recent years, interface quasi-Newton methods
have 
gained 
growing 
attention in the fluid-structure interaction 
community
by significantly improving partitioned solution schemes:
They not only help to control the inherent added-mass instability,
but also prove to substantially speed up the coupling's convergence.
In this work, we present a novel variant:
The key idea is to build on the multi-vector Jacobian update scheme first presented by Bogaers et al. \cite{Bogaers_IQNMVJ}
and avoid any explicit representation of the (inverse) Jacobian approximation,
since it slows
down the solution
for large systems.
Instead, all terms involving a quadratic complexity have been systematically eliminated.
The result is a new multi-vector interface quasi-Newton variant
whose computational cost scales linearly with the problem size.

\end{abstract}


\end{frontmatter}


\section{Introduction}

Partitioned solution schemes 
for
fluid-structure interaction (FSI)
are widely used
in modern computational engineering science,
as they offer great flexibility and modularity concerning the
two solvers employed for the fluid and the structure.
Treated as black boxes, the solvers are coupled only via the exchange of interface data.
The major downside of partitioned schemes,
however, is an inherent instability caused by the so-called
added-mass effect \cite{FoersterPhD,Wall_AddedMass}.
Depending on the application,
its influence might be severe enough to impede a numerical solution
-- if no counter-measures are taken.

The simplest way to retain stability is a relaxation of the exchanged coupling data.
The relaxation factor
might either be constant or updated dynamically via Aitken's method \cite{FSI_DynamicRelax,irons1969version}.
Unfortunately, 
the high price to be paid in terms of the coupling's convergence speed often renders this approach
infeasible.

As opposed to this,
interface quasi-Newton (IQN) methods 
have proven
capable of both stabilizing and accelerating partitioned solution schemes \cite{GatzhammerPhD,degroote2010performance,degroote2010partitioned}.
Identifying the converged solution as a fixed point,
their basic idea is to speed up the
coupling iteration
 using Newton's method.
Since the required (inverse) Jacobian
is typically not accessible,
interface quasi-Newton approaches settle for approximating it instead.

Early work in the field has been done 
by Gerbeau et al. \cite{gerbeau2003quasi} as well as van Brummelen et al. \cite{van2005interface}.
A breakthrough, however,
was the interface quasi-Newton inverse least-squares (IQN-ILS)
method by Degroote et al. \cite{PerformancePartitionedMonolithic}:
On the one hand, 
it directly solves for the inverse Jacobian required in the Newton linearization;
on the other hand, the IQN-ILS variant introduces the least-squares 
approximation based on input-output data pairs that is still common today.

Research of the last decade has shown that a reutilization of data pairs from previous time steps is extremely advantageous.
Unfortunately, an explicit incorporation in
the IQN-ILS method suffers from
numerical difficulties such as rank deficiency; moreover, good choices for the number of incorporated time steps are in general problem-dependent \cite{scheufele2017robust,ComparisonQuasiNewton}.
However, the works by Degroote and Vierendeels  \cite{Degroote_MultiSolver}  as well as Haltermann et al.  \cite{haelterman2016improving}
have shown that filtering out linearly dependent data pairs is an effective way of alleviating these issues.

%
%
%
As an alternative,
Bogaers et al. \cite{Bogaers_IQNMVJ} and Lindner et al. \cite{ComparisonQuasiNewton}
formulated a very beneficial implicit reutilization of past information,
yielding the  interface quasi-Newton inverse multi-vector Jacobian (IQN-IMVJ)
method.
Its only real drawback is the required explicit representation of the approximated inverse Jacobian:
Since the related cost is
increasing quadratically
with the number of degrees of freedom at the FSI boundary,
this explicit form seriously slows down the 
numerical simulation
for larger problem scales.

In this work, we present an enhancement of the IQN-IMVJ concept tackling exactly this shortcoming:
Retaining the effective implicit reutilization of past data,
the new interface quasi-Newton implicit multi-vector least-squares (IQN-IMVLS) method
completely avoids any explicit Jacobian or quadratic dependency 
on the interface resolution. 
While the advantages of the multi-vector approach are preserved, 
the resulting linear complexity ensures the negligibility of computational cost
even for large problem scales.
\\

This paper is structured as follows:
Section \ref{Sec:GoverningEquations} presents the governing equations of the considered
FSI problems,
before Section \ref{Sec:NumericalMethods} briefly covers the numerical methods applied to
solve them.
In Section \ref{Sec:IQN},
the IQN-ILS and the IQN-IMVJ approach are discussed in detail,
before the 
IQN-IMVLS method is
derived.
The efficiency of the new approach is validated in Section \ref{Sec:Results}
based on numerical test cases.

\section{Governing Equations} \label{Sec:GoverningEquations}

In general, fluid-structure interaction considers a fluid domain $\Omega_t^f \subset \mathbb{R}^{\nsd}$ 
and a structure $\Omega_t^s \subset \mathbb{R}^{\nsd}$,
that are connected via a shared boundary 
$\Gamma_t^{fs} = \partial \Omega_t^f \cap \partial \Omega_t^s \subset \mathbb{R}^{\nsd-1}$, the \textit{FSI interface}.
The subscript $t$ refers to the time level, while $\nsd$ denotes the number of spatial dimensions.
This section introduces the models 
and equations 
employed for the two subproblems
along with the boundary conditions interlinking their solutions
at the shared interface.

\subsection{Fluid Flow} \label{SubSec:Fluid}

The velocity $\vekt{u}^f(\vekt{x},t)$ and the pressure $p^f(\vekt{x},t)$ of 
the fluid are governed by the unsteady Navier-Stokes equations for an incompressible fluid, reading 
\begin{subequations}
\begin{alignat}{2}
	\rho^f \left( \frac{\partial \vel}{\partial t} + \vel \cdot \boldsymbol{\nabla} \vel - \vekt{f}^f \right) - \boldsymbol{\nabla}  \cdot \CauchyStress^f &= \vekt{0}	\qquad	&& \text{in} ~\Omega_t^f ~~\forall t \in [0,T] ~,\\
	\boldsymbol{\nabla}  \cdot \vel \,&= 0 && \text{in} ~\Omega_t^f ~~\forall t \in [0,T]~.
\end{alignat}
\end{subequations}
Therein, $\rho^f$ is the constant fluid density, while $\vekt{f}^f$ denotes the resultant of all external body forces
per unit mass of fluid.
For a Newtonian fluid 
with the dynamic viscosity $\mu^f$,
the Cauchy stress tensor $\CauchyStress^f$
 is modeled by Stokes law as
\begin{align}
	\CauchyStress^f( \vel, p^f) = - p^f \vekt{I} + \mu^f \left(  \nabla \vel + (\nabla \vel)^T \right)~.
\end{align}
The problem is closed by defining not only a divergence-free initial velocity field $\vekt{u}_0^f$, but also
a prescribed velocity $\vekt{g}^f$ on the Dirichlet boundary $ \Gamma_{D,\,t}^{f}$
and prescribed tractions $\vekt{h}^f$ on the
Neumann boundary $ \Gamma_{N,\,t}^{f}$ with its outer normal $\vekt{n}^f$:
\begin{subequations}
\begin{alignat}{2}
\vel (\vekt{x},t=0) =& ~\vel_0(\vekt{x}) \qquad &&\text{in} ~\Omega_0^f  ~,\\
\vel =& ~\vekt{g}^f  &&\text{on }  \Gamma_{D,\,t}^{f} ~~\forall t \in [0,T] ~, \\
\CauchyStress^f  \, \vekt{n}^f =& ~\vekt{h}^f  && \text{on }  \Gamma_{N,\,t}^{f} ~~\forall t \in [0,T] ~.
\end{alignat}
\end{subequations}

\subsection{Structural Deformation} \label{SubSec:Structure}

The response of the structure to external loads
is expressed via the displacement field $\vekt{d}^s(\vekt{x},t)$,
which is
governed by the
equation of motion,
stating
a dynamic balance of inner and outer stresses:
\begin{alignat}{2}
	\rho^s \frac{d^2 \vekt{d}^s}{dt^2} &= \boldsymbol{\nabla} \cdot \CauchyStress^s + \vekt{b}^s \qquad &&\text{in } \Omega_t^s ~~\forall t \in [0,T]~.
\end{alignat}
In this relation, $\rho^s$ denotes the material density and $\vekt{b}^s$ the resultant of all body forces per unit volume, whereas $\CauchyStress^s$ represents the Cauchy stress tensor.

As constitutive relation,
the St. Venant-Kirchhoff material model 
is used:
It relates the 2nd Piola-Kirchhoff stresses $\vekt{S} := \det (\vekt{F}) ~ \vekt{F}^{-1}  \, \vekt{T}^s  \, \vekt{F}^{-T}$
to the Green-Lagrange strains $\vekt{E} := \frac{1}{2} \left( \vekt{F}^T \vekt{F} - \vekt{I} \right)$
via a linear stress-strain law, reading
\begin{align}
\vekt{S} = \lambda^s \text{tr} \left( \vekt{E} \right) + 2 \mu^s \vekt{E} ~.
\end{align}
Therein, $\vekt{F}$ denotes the deformation gradient, while $\lambda^s$ and $\mu^s$ are the Lam\'e constants.
As the Green-Lagrange strain definition
forms a nonlinear kinematic relation, the structural model is geometrically nonlinear.
Hence, it is
capable of  representing large displacements and rotations, but only small strains \cite{ogden2001nonlinear,BatheFEM}.

Collecting all this information, the equation of motion can be expressed in the (undeformed) reference configuration $\Omega^s_0$
in a total Lagrangian fashion:
\begin{alignat}{2}
\rho^s \frac{d^2 \vekt{d}^s}{dt^2} &= \boldsymbol{\nabla}_0 \cdot \left( \vekt{S} \vekt{F}^T \right) + \vekt{b}^s \qquad &&\text{in } \Omega_0^s ~~\forall t \in [0,T]~.
\end{alignat}
Again,
the problem is closed by defining an initial displacement $\vekt{d}_0$,
which is typically zero,
and a set of boundary conditions
 on two complementary subsets of 
$\Gamma_0^s = \partial \Omega^s_0$:
Prescribing the displacement $\vekt{g}^s$ on the Dirichlet part $\Gamma_{D,0}^s$
and the
tractions $\vekt{h}^s$ on the Neumann part $\Gamma_{N,0}^s$,
with the outer normal in the reference state $\vekt{n}_0$,
the conditions read
\begin{subequations}
	\begin{alignat}{2}
	\vekt{d}^s(\vekt{x},t=0) =& ~\vekt{d}_0^s(\vekt{x} )\qquad &&\text{in } \Omega_0^s ~, \\
	\vekt{d}^s =& ~ \vekt{g}^s 		 &&\text{on }  \Gamma_{D,\,0}^{s} ~~\forall t \in [0,T] ~, \\
	\vekt{F \,S} \, \vekt{n_0} =& ~ \vekt{h}^s		 &&\text{on }  \Gamma_{N,\,0}^{s} ~~\forall t \in [0,T] ~.
	\end{alignat}
\end{subequations}

\subsection{Coupling Conditions at the Interface}

The essence of fluid-structure interaction is that the two subproblems cannot be solved independently,
since 
their solution fields are interlinked via the so-called coupling conditions arising at the
shared interface 
$\Gamma_t^{fs}$:
\begin{itemize}
	\item The kinematic coupling condition requires the continuity of displacements $\vekt{d}^f$ and $\vekt{d}^s$, velocities $\vekt{u}^f$ and $\vekt{u}^s$, and accelerations  $\vekt{a}^f$ and $\vekt{a}^s$ across the FSI boundary  \cite{NorbertPhD}:
	\begin{subequations}
			\begin{alignat}{2}
				\vekt{d}^f  &= \vekt{d}^s	\qquad &&\text{on } \Gamma_t^{fs} ~\forall t \in [0,T]~, \\
				\vel  &= \vekt{u}^s 						&&\text{on } \Gamma_t^{fs} ~\forall t \in [0,T]~, \\
				\vekt{a}^f  &= \vekt{a}^s  &&\text{on } \Gamma_t^{fs} ~\forall t \in [0,T]~. 
			\end{alignat}
	\end{subequations}
	\item In agreement with Newton's third law,
	the dynamic coupling condition,
	\begin{alignat}{2}
		\CauchyStress^f  ~ \vekt{n}^f  &= \CauchyStress^s  ~ \vekt{n}^s  	\qquad &&\text{on } \Gamma_t^{fs} ~\forall t \in [0,T]~,
	\end{alignat}	
	 \noindent enforces the equality of stresses
	 at the interface. Therein, 
	 $\vekt{n}^f$ and $\vekt{n}^s$ denote the associated normal vectors \cite{CarstenBraun_2007}.
\end{itemize}
In the continuous problem formulation,
satisfying both these coupling conditions for every time $t \in [0,T]$ ensures the conservation of mass, momentum, and energy
over the FSI boundary \cite{Kuttler_Partitioned}.

\section{Numerical Methods} \label{Sec:NumericalMethods}

\subsection{Flow Solution}
The Navier-Stokes equations formulated in Section \ref{SubSec:Fluid}
are discretized in this work using P1P1 finite elements,
i.e., linear interpolations for both velocity and pressure.
Numerical instabilities, caused by P1P1 elements violating the LBB condition,
are overcome using a \textit{Galerkin/Least-Squares (GLS)} stabilization
 \cite{LutzPhD, FEM_Flows}.

Contrary to the usual practice,
both space and time are discretized by finite elements.
More precisely,
we employ the \textit{deforming-spatial-domain/stabilized-space-time (DSD/SST)} approach introduced by  Tezduyar and Behr \cite{ArticleDSDSST_1,ArticleDSDSST_2}.
By formulating the variational form over the space-time domain,
this approach inherently accounts for an evolving spatial
domain.
While the finite-element interpolation functions are continuous in space,
the linear discontinuous Galerkin ansatz
in time allows to solve
one space-time slab after another. 

The mesh is adjusted to moving boundaries, e.g., the FSI interface, by means of interface tracking \cite{,NorbertPhD,Steffi2015}.
The resulting mesh distortion is compensated by the \textit{elastic mesh update method (EMUM)} \cite{Behr_Abraham}.

\subsection{Structural Solution}

The structural subproblem is discretized in space by \textit{isogeometric analysis} \cite{IGA_Hughes},
while the time integration is performed using the \textit{generalized-${\alpha}$} scheme \cite{Bossak_Chung,Bossak_Kuhl}.

\subsubsection{Isogeometric Analysis (IGA)}

Introduced by Hughes et al. \cite{IGA_2004} in 2005, 
isogeometric analysis is a finite-element variant aimed at achieving 
geometrical accuracy 
and a closer linkage between CAD and numerical analysis .
The essential idea is to express the solution space
via the same basis functions 
as used for the geometry description.
Since CAD systems are commonly based on non-uniform rational B-Splines (NURBS),
we choose a NURBS ansatz
for the unknown displacement field.
Mathematically, a NURBS is a linear combination of its $n$ control points $\vekt{x}_I \in \mathbb{R}^{\nsd}$
and the rational basis functions $N_{I}(\boldsymbol{\xi})$,
defined in the parameter space $\boldsymbol{\xi} \in \boldsymbol{\Xi} \subset \mathbb{R}^d$, with the spline dimension $d$ \cite{thenurbsbook}.
A NURBS surface $\vekt{\bar{x}} (\boldsymbol{\xi}) \in \mathbb{R}^{\nsd}$ hence has the form
\begin{align}
\vekt{\bar{x}}(\xi^1, \xi^2) = \sum_{I=1}^n N_I (\xi^1, \xi^2) \, \vekt{x}_I ~
\quad \text{with} ~ (\xi^1, \xi^2) = \boldsymbol{\xi}
\end{align}

\subsubsection{Shell Theory}

Shell structures are thin-walled structures capable of
providing lightweight, cost-efficient and yet stable constructions for numerous engineering applications \cite{Shells_SensitiveRelation}. 
Mathematical shell models
exploit the small thickness by reducing the structure from a volumetric description to the midsurface
plus an interpolation over the thickness $h$.
Combining isogeometric analysis with Reissner-Mindlin shell theory,
a NURBS midsurface $\vekt{\bar{x}}(\xi^1,\xi^2)$ and a linear interpolation are
chosen;
any position in the structural domain is hence expressed as
\begin{align}
\vekt{x}(\xi^1,\xi^2,\xi^3) = \vekt{\bar{x}}(\xi^1,\xi^2) + \xi^3 \vekt{b}(\xi^1,\xi^2)~.
\end{align}
The parametric coordinate $\xi^3 \in [-\frac{h}{2},+\frac{h}{2}]$ changes along the thickness direction, defined by the director vector $\vekt{b}(\xi^1,\xi^2)$.
Consequently, the unknown displacement field 
is a combination of the midsurface displacement $\vekt{\bar{d}}^s(\xi^1,\xi^2)$
and a second term accounting for changes of the director vector
$\Delta \vekt{b}(\xi^1,\xi^2)$:
\begin{align}
\vekt{d}^s(\xi^1,\xi^2,\xi^3) = 
\vekt{\bar{d}}^s(\xi^1,\xi^2) + \xi^3 \Delta \vekt{b}(\xi^1,\xi^2)~.
\end{align}

For detailed information on nonlinear isogeometric Reissner-Mindlin shell elements,
the works by Dornisch and Klinkel \cite{Dornisch_PhD,Dornisch_A,Dornisch_D}
are recommended.

\subsection{Coupling Approach}

This work pursues a partitioned FSI coupling approach:
Two distinct solvers 
are employed for the fluid and the structure;
they are
connected via a coupling module,
which
handles the exchange of interface data
 in accordance to the coupling conditions.
While the strengths of this approach are its great flexibility and modularity
regarding the single-field solvers,
these advantages come at the price of additional considerations to be made in terms of
data exchange:

\begin{itemize}
	\item 
	In general,
	 the meshes  -- or even the discretization techniques --
	do not match at the FSI boundary. Hence, a conservative projection is required to
	transfer relevant data between the two solvers.
	In this work, we employ a spline-based 
	variant of the \textit{finite interpolation elements (FIE)} method; 
	a detailed description can be found in Hosters et al. \cite{Hosters2017}.
	For the sake of observability, 
	however, any effects of this \textitbf{spatial coupling} are neglected in the following,
	since they do not interfere with the presented methods. 
	\item 
	Due to the (potentially strong) interdependency between the two subproblems,
	  a consistent FSI solution
	  in general requires an iterative procedure,
	which is referred to as \textitbf{strong (temporal) coupling} \cite{NorbertPhD,Degroote_MultiSolver}.
\end{itemize} 
	
\subsubsection{Dirichlet-Neumann Coupling Scheme}

Nowadays, the de facto standard among strong coupling approaches for FSI problems is the Dirichlet-Neumann coupling scheme
illustrated in Figure \ref{fig:DN_Coupling}:
  \begin{figure}[h!]
  	\centering
  	\resizebox{0.75\textwidth}{!}{
  		\input{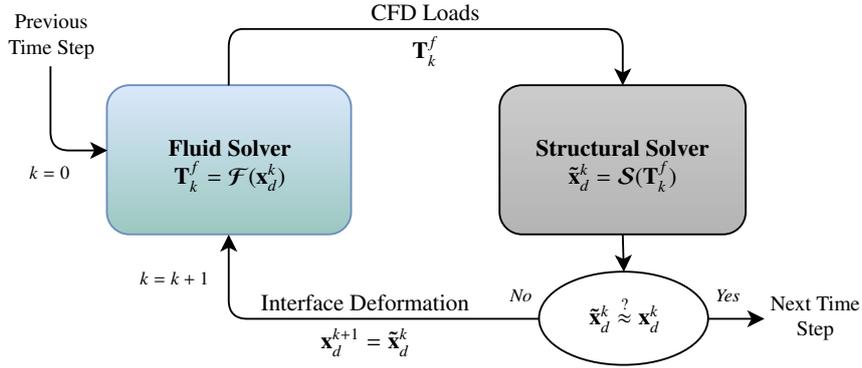}
  	}
  	\caption[Dirichlet-Neumann Scheme]{Sketch of the Dirichlet-Neumann coupling scheme.}
  	\label{fig:DN_Coupling}
  \end{figure}
The first coupling iteration ($k=0$) of a new time step $ t^{n} \to t^{n+1}$
 starts with the fluid solver: Based on the interface deformation $\vekt{x}_d^k$  
of time level $t^n$
(or an extrapolated one), it computes a new flow field. 
In compliance with
 the dynamic coupling condition,  
the resulting fluid stresses
$\FluidStressK= \fluidSolver(\vekt{x}_d^k)$ acting on $\Gamma^{fs}$ 
 are passed as a Neumann condition to the structural solver,
 which determines the corresponding interface deformation $\xtil_d^k=\structSolver(\FluidStressK)$.
If it matches (within a certain accuracy) the previous deformation , i.e., $\xtil_d^k \approx \vekt{x}_d^k$,
the coupled system has converged and we can go on to the next time step.
Otherwise, the interface deformation $\vekt{x}_d^{k+1} = \xtil_d^k$  is passed back to the flow solver as a Dirichlet condition, following kinematic continuity, and the next coupling iteration $k+1$ is started.

From a mathematical point of view, the Dirichlet-Neumann coupling scheme can be interpreted as a fixed-point iteration of the 
\textit{deformation at the interface} $\vekt{x}_d$ \cite{FSI_DynamicRelax,ComparisonQuasiNewton}; 
the fixed-point operator $\fixPntOperator(\vekt{x}_d^k) \equiv \structSolver \circ \fluidSolver(\vekt{x}_d^k)$
 corresponds to the two subsequent solver calls, i.e., to running one coupling iteration $k \to k+1$:
	\begin{align} 
	\xtil_d^k =  \fixPntOperator(\vekt{x}_d^k)~, \quad 
	\vekt{x}_d^{k+1} = \xtil_d^k \label{Eqn:Fixpoint_trivial} ~.
	\end{align}
Finding the converged solution of the next time level
is hence equivalent to finding a fixed point $\vekt{x}_d^\star = \fixPntOperator(\vekt{x}_d^\star)$ of the interface deformation.
Defining the fixed-point residual $\vekt{R}(\vekt{x}^k_d)$ and its
inverse $\Rtil(\xtil^k_d)$ as
\begin{subequations} \label{Eqn:ResidualDefinition}
\begin{alignat}{4}
\vekt{R}(\vekt{x}_d^k) &:= \fixPntOperator(\vekt{x}_d^k) ~ &-&& \vekt{x}_d^k \quad &&= \xtil_d^k - \vekt{x}_d^k ~, \\
\Rtil(\xtil_d^k) &:= \quad \xtil_d^k &-&&~ \fixPntOperator^{-1}(\xtil_d^k) &&\,= \xtil_d^k - \vekt{x}_d^k ~,
\end{alignat}
\end{subequations}
with $\Rtil(\xtil_d^k) \equiv \vekt{R}(\vekt{x}_d^k)$,
the convergence criteria can analogously be expressed as a root of the residual,
i.e., 
$\vekt{R}(\vekt{x}_d^*) = \Rtil(\vekt{x}_d^*) = \vekt{x}_d^* - \vekt{x}_d^* = \vekt{0}$.

Note that although it is much less common, the fixed-point ansatz and the convergence criteria could analogously be formulated based on the fluid loads at the FSI interface.

\subsubsection{Added-Mass Instability}

Unfortunately, partitioned algorithms 
for fluid-structure interaction involving incompressible fluids
exhibit an inherent instability,
which may be severe, depending on the simulated problem.
This so-called \textitbf{(artificial) added-mass effect} originates from
the CFD forces inevitably being determined based on a defective
structural deformation
during the iterative procedure.
%
%
%
As a consequence, they differ from the correct loads of the next time level.
This deviation
%
potentially
acts like an additional fluid mass on the structural degrees of freedom.
More precisely, due to the kinematic continuity an overestimated structural deformation
entails an exaggerated fluid acceleration 
--
 and hence excessive inertia stress terms.
In many cases the effect is amplified throughout the coupling iterations,
causing a divergence of the coupled problem. 
A major influencing factor are 
high density ratios between the fluid and the structure, $\rho^f / \rho^s$, 
but dependencies on the viscous terms and the temporal discretization are observed, too. 
We recommend 
the works by F\"orster \cite{FoersterPhD}, F\"orster et al. \cite{Wall_AddedMass}, and Causin et al. \cite{CausinAddedMass}
for more details on the added-mass effect.
Moreover, van Brummelen et al. \cite{BrummelenAddedMass} investigate similar difficulties for compressible flows.

\subsubsection{Stabilizing and Accelerating the Coupling Scheme}

One way to deal with the added-mass instabilities is to adjust
the computed interface deformation $\xtil_d^k$
by an
update step $\vekt{x}_d^{k+1} = \mathcal{U} (\xtil_d^k)$
before passing it back to the flow solver.
Depending on the chosen update technique,
both the stability and efficiency of
the coupling scheme can be improved. 

The simplest way to increase stability is a \textbfit{constant under-relaxation}
of the interface deformation:
\begin{align}
	\vekt{x}_d^{k+1} = \mathcal{U}_{Relax}(\xtil_d^k) =  \omega \xtil_d^k + (1-\omega) \vekt{x}_d^k~, \label{Eqn:Relaxation}
\end{align}
with $\omega<1$.
Effectively, it yields an
interpolation between the current and the previous interface displacements  $\xtil_d^k$ and $\vekt{x}_d^k$.
As the factor must be chosen small enough to avoid the coupling's divergence for any
time step,
unfortunately
this approach often comes at a very high price in terms of efficiency \cite{degroote2010performance}.
\textitbf{Aitken's dynamic relaxation} \cite{FSI_DynamicRelax,irons1969version} tackles this issue by 
dynamically
adapting the relaxation factor in Equation (\ref{Eqn:Relaxation}) for each coupling iteration by
\begin{align}
	\omega_k = - \omega_{k-1} \frac{ (\res{k-1})^T (\res{k} - \res{k-1} ) }{ \| \res{k} - \res{k-1} \|_2^2 }~,
\end{align}
i.e., based on the two most recent fixed-point residuals $\res{k}$ and $\res{k-1}$.
Despite its rather simple implementation,
in many cases
Aitken's relaxation provides significant speed-ups 
without perturbing the stability of the relaxation.
Still, the performance of the coupling scheme can be pushed further by 
more sophisticated approaches like
\textit{interface quasi-Newton (IQN)} methods. \\

\mypara
In the following,
we will express the interface deformation  
as the structural solution at the FSI interface.
While its representation via the boundary displacement of the fluid mesh is likewise possible,
the typically finer interface resolution would result in higher numerical cost
for the discussed update techniques.

\section{Interface Quasi-Newton Methods} \label{Sec:IQN}

\subsection{Interface Quasi-Newton Inverse Least-Squares (IQN-ILS) Method} \label{SubSec:IQN-ILS}

In a partitioned solution procedure for fluid-structure interaction, 
the single-field solvers
are typically the most expensive part,
rendering all further costs, e.g., 
for data exchange,
negligible in comparison \cite{Bogaers_IQNMVJ,uekermann2016partitioned}.
Since
the converged time step solution 
 has been identified as a root of the
fixed-point residual $\vekt{R}(\vekt{x}_d)$ or its inverse form $\Rtil(\xtil_d)$
defined in Equation (\ref{Eqn:ResidualDefinition}),
increasing the efficiency essentially comes down to reducing the number of coupling iterations required
to find such a root.
Consequently,
employing Newton's method as an update step of the interface deformation
would be a promising approach to speed up convergence, yielding \cite{PerformancePartitionedMonolithic,ComparisonQuasiNewton}
\begin{align}
\vekt{x}^{k+1}_d =  \mathcal{U}_{Newton}(\xtil_d^k) = \vekt{\tilde{x}}^k_d - \vekt{J}^{-1}_{\tilde{R}} (\vekt{\tilde{x}}^k_d)~\vekt{\tilde{R}}(\vekt{\tilde{x}}^k_d) ~.
\label{Eqn:NewtonUpdate}
\end{align}
Unfortunately, an evaluation of the required inverse Jacobian $\vekt{J}^{-1}_{\tilde{R}}:=  d\Rtil / d\xtil_d$ 
in general is not accessible, as it would involve the derivatives of 
both the fluid and the structural solver.  
The idea of \textitbf{interface quasi-Newton} methods is to circumvent this issue by approximating the (inverse) Jacobian rather than evaluating it exactly.
Using a Taylor expansion of $\Rtil(\xtil_d)$, 
we can approximate $\vekt{J}^{-1}_{\tilde{R}}$ via
\begin{align}
	\vekt{J}_{\tilde{R}}^{-1} (\xtil_d^k) ~  \Delta \vekt{R} (\Delta \xtil_d) \approx  \Delta \xtil_d ~,
\end{align}
based on an input-output data pair.
Such a pair is formed by some change in the interface deformation $\Delta \xtil_d \in \mathbb{R}^m$
and the corresponding change of the fixed-point residual $\Delta \vekt{R} \in \mathbb{R}^m$.
Therein, $m$ denotes the number of structural degrees of freedom \textit{at the FSI interface}.
Essentially, this idea is an $m$-dimensional version of 
approximating a derivative with the slope of a secant.
To avoid additional solver calls,
the required data pairs are formed from
the intermediate results $\xtil_d^i$ and $\Rtil^i$
of the
$k$ coupling iterations already performed for the current time step.
More precisely, they are stored in the
\textit{input matrix} $\vekt{V}_k \in \mathbb{R}^{m \times k}$
and the \textit{output matrix} $\vekt{W}_k \in \mathbb{R}^{m \times k}$ \cite{scheufele2017robust}:
\begin{subequations} \label{Eqn:InputOutputMatrices}
	\begin{alignat}{7}
	\vekt{V}_k &= \big[ \Delta \vekt{R}_0^1, &&\Delta \vekt{R}_1^2,~ && \cdots, ~ &&\Delta \vekt{R}_{k-1}^k && \big] 
	&&\text{with }
	\Delta \vekt{R}_i^j &&=  \Rtil(\xtil_d^j) - \Rtil(\xtil_d^i)  = \vekt{R}(\vekt{x}_d^j) - \vekt{R}(\vekt{x}_d^i)   ~, \\
	\vekt{W}_k &= \big[ \Delta \vekt{\tilde{x}}_0^1, ~ &&\Delta \vekt{\tilde{x}}_1^2,~ && \cdots, && \Delta \vekt{\tilde{x}}_{k-1}^k   && \big] 
\qquad 
&& \text{with }
\Delta \vekt{\tilde{x}}_i^j &&= \vekt{\tilde{x}}_d^j - \vekt{\tilde{x}}_d^i ~.
	\end{alignat}
\end{subequations}
With  the collected data, an approximation
 $\Jac \approx \vekt{J}^{-1}_{\tilde{R}} \in \mathbb{R}^{m \times m}$
 of the inverse Jacobian
  can be formulated via
\begin{align}
\Jac ~ \vekt{V}_k = \vekt{W}_k ~. \label{Eqn:LinSysEq1}
\end{align}
Since the number of data pairs 
stored in $\vekt{W}_k$ and $\vekt{V}_k$
typically is much smaller than 
the number of structural
degrees of freedom at the FSI interface, 
i.e., $k << m$,
the linear system of equations (\ref{Eqn:LinSysEq1})
is underdetermined. 
The existence of a unique solution is ensured by demanding the minimization of the Frobenius norm
 \cite{ComparisonQuasiNewton}:
\begin{align}
\|\, \Jac  \|_F \to \min \label{Eqn:NormMin} ~.
\end{align}
Together, equations (\ref{Eqn:LinSysEq1}) and (\ref{Eqn:NormMin}) form a constrained optimization problem, which
leads to an explicit form of the inverse Jacobian approximation \cite{scheufele2017robust,ComparisonQuasiNewton}:
\begin{align}
\Jac  = \vekt{W}_k \left( \vekt{V}_k^T \vekt{V}_k \right) ^{-1} \vekt{V}_k^T ~. \label{Eqn:JacobApprox}
\end{align}
Inserting this inverse Jacobian approximation into the Newton update formula in Equation (\ref{Eqn:NewtonUpdate})
yields a quasi-Newton update scheme for the interface deformation, 
\begin{align}
\vekt{x}_d^{k+1} &= \mathcal{U}_{IQN}(\xtil_d^k) =
 \xtil_d^k +\Jac \left( - \vekt{R}(\vekt{x}_d^k) \right) =
\xtil_d^k + \vekt{W}_k  \underbrace{  \left( \vekt{V}_k^T \vekt{V}_k \right) ^{-1} \vekt{V}_k^T   ~\left( - \vekt{R}(\vekt{x}_d^k) \right)  }_{:=  \boldsymbol{\alpha}}\label{Eqn:IQN_ILS_Update} 
= \xtil_d^k + \vekt{W}_k \boldsymbol{\alpha}
~. 
\end{align}
Defining the vector $\boldsymbol{\alpha} = \left( \vekt{V}_k^T \vekt{V}_k \right) ^{-1} \vekt{V}_k^T   ~\left( - \vekt{R}(\vekt{x}_d^k) \right)$
exploits that 
the inverse Jacobian is not needed explicitly here, but only its product with the residual vector.
An efficient way of computing $\boldsymbol{\alpha} \in \mathbb{R}^{k}$ is to solve the least-squares problem \cite{PerformancePartitionedMonolithic,scheufele2017robust}
\begin{align} 
 \min_{\boldsymbol{\alpha} \in \mathbb{R}^k} \| \vekt{V}_k \boldsymbol{\alpha} + \vekt{R}(\vekt{x}_d^k)\|_2 ~,  \label{Eqn:LSalpha}
\end{align}
e.g., using a QR decomposition via Householder reflections.
In the first coupling iteration, a relaxation step is used, as no data pairs are available yet.

The approach presented so far is referred to as \textitbf{interface quasi-Newton inverse least-squares (IQN-ILS)} method \cite{PerformancePartitionedMonolithic};
it can be interpreted as the basis of the other
IQN variants
discussed in this work.

\subsection{Benefit from Previous Time Steps} \label{SubSec:BenefitFromPastData}

So far, the inverse Jacobian of the residual operator is approximated 
only based on information gathered in the coupling iterations of the current time step.
In principle, however, the efficiency of IQN methods can be significantly increased by incorporating data
from previous time steps as well.
The most straightforward way to do so is to explicitly include the data pairs of past time steps in the input and output matrices $\Vk$ and $\Wk$ \cite{PerformancePartitionedMonolithic}.

Unfortunately, 
apart from increasing costs for handling big data sets,
a growing number of data pairs stored in these matrices entails two major problems \cite{Bogaers_IQNMVJ,ComparisonQuasiNewton,IQN_POD}:
	(1) The matrix $\Vk$ might be very close to rank-deficient due to (almost) linear dependent columns;
	related to that, the condition number of the least-squares problem quickly increases;
	(2) information gathered in different time steps might be contradictory.
Together, these drawbacks carry the risk of preventing the system from being numerically solvable at all.

An obvious remedy is to incorporate only the data
of the $q$ most recent time steps.
While this approach is in fact capable of yielding a superior convergence speed,
its major shortcomings are 
the risk of rank deficiency and 
in a good choice for the parameter $q$ being problem-dependent
\cite{Bogaers_IQNMVJ,scheufele2017robust}.
One way to mitigate these drawbacks
is by employing a filtering technique \cite{Degroote_MultiSolver,haelterman2016improving}.

\subsection{Interface Quasi-Newton Inverse Multi-Vector Jacobian Method} \label{SubSec:IQN_IMVJ}

An alternative way of reusing data from previous time steps was introduced by Bogaers et al. \cite{Bogaers_IQNMVJ}
and developed further by Lindner et al. \cite{ComparisonQuasiNewton}:
The \textitbf{interface quasi-Newton inverse multi-vector Jacobian (IQN-IMVJ)} method
 combines the 
IQN
 approach with the idea of Broyden's method:
 In a Newton iteration, one might expect the Jacobians evaluated in subsequent iterations to be similar in some sense.
 Therefore, Broyden's method limits the typically costly
 Jacobian evaluation to the first iteration only;
 after that, the new Jacobian is instead
 approximated by a
 rank-one update of the previous one.
For the interface quasi-Newton framework, this concept is adopted by formulating the
current inverse Jacobian approximation $\JacN{n+1}$ 
as an update of the one determined in the previous time step $\JacN{n}$:
\begin{align}
	\JacN{n+1} = \JacN{n} + \Delta \Jac ~, 	\label{Eqn:BasicJacUpdate}
\end{align}
where $\Delta \Jac$ denotes the update increment. 
Introducing this concept changes the
constrained optimization problem 
discussed in Section \ref{SubSec:IQN-ILS} to
\begin{align}
	\| \Delta \Jac \|_F =  \| \JacN{n+1} - \JacN{n} \|_F \to \min  \qquad \text{subject to} \qquad  \Delta \Jac ~\Vk = \Wk - \JacN{n} \Vk~.
\end{align}
This system 
allows for a quite descriptive interpretation:
Fitting Broyden's idea, the difference between the inverse Jacobians
of the current and the previous time step is minimized,
rather than the norm of the approximation itself.

Again, the result is an explicit approximation of the current inverse Jacobian \cite{ComparisonQuasiNewton}:
\begin{align}
	\JacN{n+1} = \JacN{n}  + \left( \Wk - \JacN{n}  \, \Vk \right)  \underbrace{ \left( \Vk^T \Vk \right)^{-1} \Vk^T }_{:= \vekt{Z}_k} = \JacN{n} + \left( \Wk - \JacN{n}  \, \Vk \right) \vekt{Z}_k~. \label{Eqn:IMVJJacUpdate}
\end{align}
In contrast to the IQN-ILS approach, the IQN-IMVJ method
explicitly updates the Jacobian in every coupling iteration via Equation (\ref{Eqn:IMVJJacUpdate}).
Therein, the matrix $\vekt{Z}_k = \left( \Vk^T \Vk \right)^{-1} \Vk^T \in \mathbb{R}^{k\times m}$
is determined by 
solving the least-squares problem 
\begin{align}
\min_{\vekt{z}_j \in \mathbb{R}^{k} }  \| \vekt{V}_k \vekt{z}_j - \hat{\vekt{e}}_j \|_2		\label{LS_Z}
\end{align}
for each column $\vekt{z}_j$ with the $j$-th unit vector  $\hat{\vekt{e}}_j $,
again using a Householder QR decomposition.
After 
$\JacN{n+1}$
has been updated,
the quasi-Newton step is performed:
\begin{align}
\vekt{x}_d^{k+1} =\mathcal{U}_{IQN}(\xtil_d^k) = \xtil_d^k - \JacN{n+1} \vekt{R}^k ~, \qquad \text{where } \vekt{R}^k := \vekt{R}(\vekt{x}_d^k)
\end{align}
Since the inverse Jacobian 
 is initialized to zero, i.e., $\JacN{0} =\vekt{0}$, 
the first time step is equivalent to the IQN-ILS method.

Note that the matrices $\Wk$ and $\Vk$ only contain data collected in the current time step;
because, rather than 
explicitly including data pairs collected in previous time steps, the
IQN-IMVJ approach incorporates them implicitly
in  form of the previous Jacobian approximation $\JacN{n} $.
This implicit reutilization of past information entails some significant advantages \cite{Bogaers_IQNMVJ,ComparisonQuasiNewton}:
(1) Since the matrices $\Wk$ and $\Vk$ are not affected,
	neither is the condition number of the least-squares problem;
(2) it renders any tuning of (strongly) problem-dependent parameters obsolete;
(3) since past information is matched in a minimum norm sense only, it is automatically less emphasized than that of the current time step. This avoids the risk of relying on outdated or contradicting data.

These benefits 
come at the price of
an explicit representation of the inverse Jacobian approximation, that is often very expensive to store and
handle, as the entailed complexity is growing quadratically with the problem scale.

\subsection{Interface Quasi-Newton Implicit Multi-Vector Least-Squares Method} \label{SubSec:IQN-IMVLS}

In this paper, we introduce an enhancement of the IQN-IMVJ approach
tackling its main issue, i.e., the high cost
related to the explicit Jacobian approximation.
By successively replacing the expensive parts, 
this section derives the new \textitbf{interface quasi-Newton implicit multi-vector least-squares (IQN-IMVLS)} method. \\

As a first step,
the explicit use of the current inverse Jacobian approximation $\JacN{n+1}$ is eliminated:
By inserting the Jacobian update (\ref{Eqn:IMVJJacUpdate}) into the quasi-Newton step,
the vector $\boldsymbol{\alpha}=\left( \vekt{V}_k^T \vekt{V}_k \right)^{-1} \vekt{V}_k^T \, (-\vekt{R}^k)$
can be identified in analogy to the IQN-ILS approach,
changing the update formula to
\begin{align}
	\vekt{x}^{k+1}_d
	= \xtil_d^k - \JacN{n+1} \vekt{R}^k 
	=  \xtil_d^k - \JacN{n} \vekt{R}^k + \left( \Wk - \JacN{n} \, \Vk \right)  \underbrace{ \left( \Vk^T \Vk \right)^{-1} \Vk^T (- \vekt{R}^k) }_{:=\boldsymbol{\alpha}} 
	=
 \xtil_d^k - \JacN{n} \vekt{R}^k + \left( \vekt{W}_k - \JacN{n} \vekt{V}_k \right) \boldsymbol{\alpha} ~. \label{Eqn:IMVLS_implicitUpdate} 
\end{align}
By reducing the least-squares problem back to Equation (\ref{Eqn:LSalpha}),
this drastically reduces the computational cost.
The past inverse Jacobian $\JacN{n} \in \mathbb{R}^{m \times m}$, however, is still explicitly required:
Once a time step has converged, 
it is updated to the next time level by Equation (\ref{Eqn:IMVJJacUpdate});
the associated cost is quadratic in $m$.
Beyond that,
$\JacN{n}$ is involved in the quasi-Newton formula in Equation (\ref{Eqn:IMVLS_implicitUpdate}). In particular, it is needed for
the potentially very costly matrix product
 in $\vekt{B}_{k} := (\Wk - \JacN{n} \Vk)$,
which has a complexity of $\mathcal{O}(m^2 \, k)$.
As mentioned before, however,
the matrices $\vekt{W}_k$ and $\vekt{V}_k$ 
are successively built up 
only by appending new columns.
Taking this into account, we can reformulate
\begin{align}
\vekt{B}_{k} =
\Wk - \JacN{n} \Vk &= \Big[ \vekt{W}_{k-1}, ~~ \Delta \xtil^k_{k-1} \Big] - \JacN{n} \left[ \, \vekt{V}_{k-1} , ~~ \vekt{R}^k - \vekt{R}^{k-1} \right] 
= \Big[ \, \underbrace{ \vekt{W}_{k-1} - \JacN{n} \vekt{V}_{k-1} }_{:= \vekt{B}_{k-1}},   ~~   \Delta \xtil^k_{k-1} -  \underbrace{ \JacN{n} \vekt{R}^k}_{:= \vekt{b}^{k}} + \underbrace{ \JacN{n} \vekt{R}^{k-1}}_{:=\vekt{b}^{k-1}} \Big] ~. \nonumber
\\ &= \left[ \, \vekt{B}_{k-1}, \quad \Delta \xtil^k_{k-1}  - \vekt{b}^{k}  + \vekt{b}^{k-1} \right] 
\end{align}
As a consequence,
restoring the terms 
$\vekt{B}_{k-1}:= \vekt{W}_{k-1} -  \JacN{n} \,\vekt{V}_{k-1} \in \mathbb{R}^{m \times k}$ 
and $\vekt{b}^{k-1} := \JacN{n} \,\vekt{R}^{k-1} \in \mathbb{R}^{m}$ from the previous iteration
in fact allows 
to reduce the number of matrix-vector products with the inverse Jacobian to one per coupling iteration, that is $\vekt{b}^k =\JacN{n} \, \vekt{R}^k$. 
This Jacobian-vector product 
remains the only operation
in Equation 
(\ref{Eqn:IMVLS_implicitUpdate})
involving an $\mathcal{O}(m^2)$ complexity.
For an increasing number of structural degrees of freedom at the interface $m$, this term's cost therefore may become dominant
and slow down the overall procedure (see Remark \ref{Remark3}).

In order to tackle this issue, we introduce an alternative purely implicit formula 
for evaluating $\vekt{b}^k = \JacN{n} \, \vekt{R}_k$ without any explicit previous Jacobian $\JacN{n}$.
Note that the explicit Jacobian update will become obsolete as a direct consequence.
Unraveling the recursion
in Equation (\ref{Eqn:IMVJJacUpdate}),
the update of the inverse Jacobian can be reformulated to 
\begin{align}
\JacN{n+1} =  \JacN{0} \prod_{i=0}^n \left( \vekt{I} - \vekt{V}_k^i \vekt{Z}_k^i \right) + \sum_{i=0}^n \vekt{W}_k^i \vekt{Z}_k^i  \prod_{j=i+1}^n \left( \vekt{I} - \vekt{V}_k^j \vekt{Z}_k^j \right) ~,
\end{align}
see \ref{AppendixProofA} for the proof.
Therein, the upper index of the matrices $\Wk^i$, $\Vk^i$, and $\Zk^i$ refers to the time step $t^{i} \to t^{i+1}$
they were determined in.
Considering the initial Jacobian approximation, $\JacN{0}=\vekt{0}$, this expression simplifies to
\begin{align}
\JacN{n+1} =  \sum_{i=0}^n \vekt{W}_k^i \vekt{Z}_k^i  \prod_{j=i+1}^n \left( \vekt{I} - \vekt{V}_k^j \vekt{Z}_k^j \right) ~.
\end{align}
While being mathematically equivalent, this formulation entails two main advantages:
First, the matrix-vector product $\JacN{n} \, \vekt{R}^k$ can be determined in an implicit manner via
\begin{align}
\JacN{n} \vekt{R}^k =  \sum_{i=0}^{n-1} \vekt{W}_k^i \vekt{Z}_k^i  \prod_{j=i+1}^{n-1} \left( \vekt{R}^k - \vekt{V}_k^j \vekt{Z}_k^j  \vekt{R}^k \right) ~,		\label{Eqn:ImplictJacRn}
\end{align}
without an $\mathcal{O}(m^2)$ complexity, and hence potentially cheaper. 
However, the cost for evaluating the expression in Equation (\ref{Eqn:ImplictJacRn}) obviously increases with the 
number of processed time steps. 
This issue is addressed using the second advantage:
While data from past time steps is still incorporated in an implicit manner, with all the associated advantages discussed in Section \ref{SubSec:IQN_IMVJ}, 
the new formulation explicitly identifies the contribution of each time step,
in the form of
the matrices 
$\Vk^i$, $\Wk^i$, and $\vekt{Z}_k^i$.
As a consequence, this update technique allows to incorporate only the $q$ most recent time steps for the previous Jacobian approximation as a way of limiting the cost of the matrix-vector product,
which then reads
\begin{align}
\JacN{n} \vekt{R}^k \approx  \sum_{i=n-q}^{n-1} \vekt{W}_k^i \vekt{Z}_k^i  \prod_{j=i+1}^{n-1} \left( \vekt{R}^k - \vekt{V}_k^j \vekt{Z}_k^j  \vekt{R}^k \right) ~.
\end{align}
In fact, by taking advantage of the repetition of terms in the product (see \ref{Appendix:JacobianProduct}),
this expression can be evaluated in $\mathcal{O}(m \, \bar{k} \, q)$,
where $\bar{k}$ denotes the average number of coupling iterations per time step.
This step can be justified by the fact that a certain time step's contribution to the previous Jacobian approximation is
gradually becoming less and less important
the further the simulation progresses.

At a first glance, one might argue that the parameter $q$ 
reintroduces exactly the problems
the multi-vector approach was designed to avoid in the first place, i.e., the dependency of the 
IQN-ILS method on the number of reused time steps.
However, this issue arose from the explicit incorporation of past data, e.g., due to high condition numbers and linear-dependent columns in $\Vk$.
Since the IQN-IMVLS approach reuses data in an implicit manner,
it does not suffer from these drawbacks.
Instead, the method's quality
in general benefits from increasing the number of past time steps taken into account,
as the limit of reusing all steps 
is analytically equivalent to the explicit Jacobian multiplication.

For this variant, the matrix $\vekt{Z}_k = \left( \Vk^T \Vk \right)^{-1} \Vk^T \in \mathbb{R}^{k\times m}$ 
still has to be determined and stored once after each time step. Since the complexity of computing it via the $m$ least-squares 
problems (\ref{LS_Z}) is $\mathcal{O}(m^2 \bar{k}^2)$ for the Householder QR approach,
i.e., growing quadratically with $m$,
we use a matrix inversion of $\Vk^T \Vk$  via a LU decomposition using partial pivoting with row interchanges instead \cite{golub1996matrix}.
While being slightly less robust to bad conditioning,
the big advantage is that 
it requires $\mathcal{O}(m \, \bar{k}^3)$ operations
and therefore scales linearly with the problem size.

Combining all this,
the modified multi-vector update completely avoids
any $\mathcal{O}(m^2)$-terms that might slow down the IQN-IMVLS approach for large systems,
where $m >> \bar{k} \, q$ holds, see Table \ref{Tab:IMVLS_complexity}. 
A pseudo-code realization of the purely implicit IQN-IMVLS method is
outlined in Algorithm \ref{Alg:IQN-IMVLS}.

\begin{table}[h]
	\centering
	\begin{tabular}{ | c | c | c | c  |}
				\hline
		\multicolumn{4}{| c | } {  \multirow{2}{*} {\textitbf{Increment step:} $\Delta \vekt{x} = \left( \Wk - \JacN{n} \, \Vk \right) \boldsymbol{\alpha}$ 
			}} \\
			\multicolumn{4} {|c |} { }  \\
			\hline  \hline
			{\textbf{Operation}}& {\textbf{Expression}} & \textbf{{Explicit}} &\textbf{{Implicit}}  \\
			\hline 	\hline
			{Least-squares problem} &    \multirow{2}{*}{$\boldsymbol{\alpha}$ }  & \multirow{2}{*}{$m\, \bar{k}^2$} & \multirow{2}{*}{$m\, \bar{k}^2$} \\
			{via Householder QR}    & &  &
			\\
			\hline
			{Matrix-vector product} \parbox[][0.65cm]{0pt}& $\vekt{b}^k = \JacN{n} \, \vekt{R}_k$ & $m^2$   & $\boldsymbol{m \, \bar{k} \, q}$ \\[1pt]
			\hline
				{Compute new column} \parbox[][0.65cm]{0pt} & $\vekt{B}_k = \left[ \, \vekt{B}_{k-1}, ~~ \Delta \xtil^k_{k-1}  - \vekt{b}^{k}  + \vekt{b}^{k-1} \right] $ 	& $m$  & $m$ \\
			\hline
			{Matrix-vector product} \parbox[][0.65cm]{0pt} & $\vekt{B}_k ~\boldsymbol{\alpha}$ 	& $m \, \bar{k}$ & $m \, \bar{k}$ \\
			\hline
			\multicolumn{4}{| c | } {  \multirow{2}{*} {\textitbf{Jacobian update: }  $\JacN{n+1} = \left( \Wk - \JacN{n} \, \Vk \right) \vekt{Z}_k = \vekt{B}_k \, \vekt{Z}_k$ 
				} } \\
				\multicolumn{4} {|c |} { }  \\
				\hline \hline
				\textbf{Operation}& \textbf{Expression} & \textbf{Explicit} & \textbf{Implicit} \\
				\hline  \hline
				{Least-squares problem} &    \multirow{2}{*}{$\vekt{Z}_k$ }  & \multirow{2}{*}{$m^2\, \bar{k}^2$} & \multirow{2}{*}{ \textbf{--} }\\
				{via Householder QR}   & & & \\
				\hline
			{Matrix-matrix product} \parbox[][0.65cm]{0pt} & $\vekt{B}_k \, \vekt{Z}_k$ & $m^2 \, \bar{k}$ & \textbf{--}\\
				\hline
				\multicolumn{4}{| c | } {  \multirow{2}{*} {\textitbf{Determine and store $\vekt{Z}_k$}  	} } \\
								\multicolumn{4} {|c |} { }  \\
				\hline \hline
				\textbf{Operation}& \textbf{Expression} & \textbf{Explicit} & \textbf{Implicit} \\
				\hline  \hline
								{Matrix inversion via} &    \multirow{2}{*}{$\left( \Vk^T \Vk \right)^{-1}$ }  & \multirow{2}{*}{ \textbf{--} }& \multirow{2}{*}{$m \, \bar{k}^3$}  \\
				{LU decomposition}   & & & \\
				\hline
				{Matrix-matrix product} \parbox[][0.6cm]{0pt}& $\left( \Vk^T \Vk \right)^{-1} \Vk^T $  & \textbf{--} & $m \, \bar{k}^2$ \\
				\hline
			\end{tabular}
			\caption[] {Complexities of the suboperations required for the IQN-IMVLS update method with an explicit previous Jacobian compared to the purely implicit version.}
			\label{Tab:IMVLS_complexity}
		\end{table}
In direct analogy to the computational complexity, the memory requirements, too, are no longer scaling quadratically but linearly with the problem size:
Although the matrices 
$\Vk^i, \Wk^i \in \mathbb{R}^{m \times \bar{k}}$ as well as $\vekt{Z}_k^i \in \mathbb{R}^{\bar{k} \times m}$
have to be stored for the $q$ most recent time steps,
the required amount of storage is much smaller than for the explicit Jacobian as long as $m >> \bar{k} q$ holds.

Of course, the effectiveness of the purely implicit IQN-IMVLS update strongly depends on this ratio of $m$ and $\bar{k} \, q$,
so that the explicit Jacobian approximation might still be the better option for small systems.
However, the assumption $m>>\bar{k} \, q$ is not very restrictive for common application scales. \\
\ \\

\begin{algorithm}[h!]
	\Vhrulefill \\
	
	\comment{\textbf{Time Step Loop:}} \For{$n=0,\cdots$}
	{
		\begin{tabularx} {\textwidth}{ @{}l X}
			\comment{Initialization:} & $\vekt{W}_0^n = [ \, ]$, $\vekt{V}_0^n = [ \,]$, $\vekt{B}_0 = [ \, ]$  \\
			\comment{First iteration:} & $\xtil_d^0 = \fixPntOperator(\vekt{x}_d^0) $ \\
			\comment{Form residual:} &  $\vekt{R}^0 = \xtil_d^0 - \vekt{x}_d^0$ \vspace{0.0cm} \\
		\end{tabularx}
		%
		\eIf{$n==0$}
		{
			\begin{tabularx} {\textwidth}{ @{}l X}
				\comment{Under-relaxtion step:} & $\vekt{x}_d^1 = \omega \xtil_d^0 + (1-\omega)~ \vekt{x}_d^0$ 
			\end{tabularx}
		}
		{
			\begin{tabularx} {\textwidth}{ @{}l X}
				\vspace{0.1cm}
				\comment{Previous inverse Jacobian times residual (implicitly):} & $\vekt{b}^0 = \sum_{i=n-q}^{n-1} \vekt{W}_k^i \vekt{Z}_k^i  \prod_{j=i+1}^{n-1} \left( \vekt{R}^0 - \vekt{V}_k^j \vekt{Z}_k^j  \vekt{R}^0 \right)$ \\
				\comment{IQN-update with Jacobian product:} & $\vekt{x}_d^1  = \xtil_d^0 - \vekt{b}^0$ 
			\end{tabularx}
		}
		\comment{\textbf{Coupling Loop:}} \For{$k=1,\cdots$ \text{until convergence}}
		{ 
			\begin{tabularx} {\textwidth}{ @{}l X}
				\comment{Perform coupling iteration:} &  $\xtil_d^k ~~~~= \fixPntOperator(\vekt{x}_d^k)$ \\
				\comment{Form residual:} & $\vekt{R}^k ~~~= \xtil_d^k - \vekt{x}_d^k$ \\
				\comment{Append new column to input matrix:} &  $\vekt{V}_k^n ~= [ \vekt{V}_{k-1}^n, ~ \, \Delta \vekt{R}_{k-1}^k ] 
				\quad \text{with}~ \Delta \vekt{R}_i^j = \vekt{R}^j - \vekt{R}^i$  \\
				\comment{Append new column to output matrix:} &  $\vekt{W}_k^n = [ \vekt{W}_{k-1}^n, ~ \Delta \xtil_{k-1}^k] 
				\quad \text{with}~ \Delta \xtil_i^j ~= \xtil_d^j - \xtil_d^i$ \\
				\comment{Previous inverse Jacobian times residual (implicitly):}  & $\vekt{b}^k ~\, = \sum_{i=n-q}^{n-1} \vekt{W}_k^i \vekt{Z}_k^i  \prod_{j=i+1}^{n-1} \left( \vekt{R}^k - \vekt{V}_k^j \vekt{Z}_k^j  \vekt{R}^k \right)$ \\
				\comment{Build $\vekt{B}_k$ restoring terms from previous iteration:} & $\vekt{B}_k ~ = [ \vekt{B}_{k-1}, ~~ \Delta \xtil_{k-1}^k + \vekt{b}^{k-1} - \vekt{b}^{k} ]$ \\ 
				\comment{Solve least-squares problem:} &  $\min \| \vekt{V}_k^n \boldsymbol{\alpha} + \vekt{R}^k \|_2$ \\
				\comment{Implicit update step:} & $\vekt{x}_d^{k+1} ~= \xtil_d^k - \vekt{B}_k \, \boldsymbol{\alpha}$ 
			\end{tabularx}
		}
		\begin{tabularx} {\textwidth}{ @{}l X}
			\comment{Determine $\vekt{Z}_k^n$ via LU decomposition:} &  $\vekt{Z}_k^n = \left( \Vk^T \Vk \right)^{-1} \Vk^T$ \\
			\comment{Store for future time steps:} & $\vekt{Z}_k^n$, $\vekt{W}_k^n$, $\vekt{V}_k^n ~\to$ Store  
		\end{tabularx} 
	}
	\Vhrulefill \\
%
	\caption[]{\footnotesize{Pseudo-code of the interface quasi-Newton implicit multi-vector least-squares (IQN-IMVLS) method.}}
	\label{Alg:IQN-IMVLS}
\end{algorithm}

\mypara 
Aside from suggesting improvements
similar to using an explicit past Jacobian within the IQN-IMVLS method,
Scheufele and Mehl \cite{scheufele2017robust} also derived a multi-vector variant with linear complexity.
Therein, the past inverse Jacobian is represented based
on the matrices $\vekt{Z}_k^i$ and $\vekt{B}_k^i$ from past time steps, i.e.,
\begin{align}
\JacN{n} =  
\sum_{i=0}^{n-1} \vekt{B}_k^i \vekt{Z}_k^i ~. \label{Eqn:IQN-RS-SVD_Sum}
\end{align}
While the idea to avoid the costly explicit Jacobian via a sum formulation is similar to the IQN-IMVLS method,
there is one essential difference:
Due to the recursive definition of the multi-vector Jacobian approximation,
each matrix $\vekt{B}_k^i =  \Wk^i - \JacN{i} \Vk^i$ contains information from the first $i$ time steps.
By implication, this means 
that the contribution of a time step $j$ is contained in all subsequent matrices 
$\vekt{B}_k^j$, $\vekt{B}_k^{j+1}$,  $\vekt{B}_k^{j+2}$, and so forth.
As a consequence, it is impossible to drop old time steps 
from the approximation while keeping more recent ones,
as the IQN-IMVLS method does.

Instead, Scheufele and Mehl
divide the simulation into chunks of several time steps;
after each of these chunks, the sum in equation (\ref{Eqn:IQN-RS-SVD_Sum}) has to be reset.
Since a plain erasure of old data at every restart dramatically impairs the
efficiency of the multi-vector approach,
focus is put on different restart options:
In particular, they introduce the \bolditalic{multi-vector Jacobian restart singular value decomposition (IQN-MVJ-RS-SVD)} method,
in which a truncated SVD accounts for the dropped data.
It introduces two main parameters:
While the influence of the chunk size $M$ is typically less marked,
the method's accuracy and efficiency strongly depends on the threshold $\varepsilon_{svd}$ for cutting off insignificant singular values.
%
However, choices around $\varepsilon_{svd} \approx 0.01$ were shown to be suitable for various test cases \cite{scheufele2017robust}.
For an in-depth discussion of the method refer to Scheufele and Mehl \cite{scheufele2017robust,ScheufelePhD}.

In contrast to this restart-based approach,
the IQN-IMVLS method
always considers the $q$ most recent time steps.
In combination with its arguably simpler implementation,
it 
therefore represents a straightforward 
way of  achieving a linear complexity
without any need for 
restart techniques. \\

\mypara \label{Remark2} 
In the beginning of a new time step,
the input and output matrix
of 
the IQN-IMVLS method
rely on very few data pairs only.
To improve the significance of the least-squares problem, 
it can be beneficial to include the information
of the most recent time step explicitly in the matrices $\Vk$ and $\Wk$ -- in addition to the implicit
reutilization of past data.
The effect of this option will be included in the  discussion of the numerical test cases in Section \ref{Sec:Results}. \\

\mypara \label{Remark3} 
The computational cost of the interface quasi-Newton approaches discussed in this work
increases with the number of structural degrees of freedom at the FSI interface.
To assess the significance of this cost in comparison to the solver calls,
the complexity of the structural solution is therefore of particular interest.
Nowadays, the structural subproblem is typically solved using some finite-element variant,
such as isogeometric analysis.
These procedures can be subdivided into two main tasks:
(1) The assembly of the system matrices is done in an element-by-element manner and hence
scales with the number of elements $n_{el}$, i.e.,
$\mathcal{O}(n_{el})$. Moreover, each of the $n_{dof}$ degrees of freedom causes
$b$ nonzero entries in the sparse system matrix \cite{farmaga2011evaluation}.
As this number $b<<n_{dof}$ primarily depends on the element type
this yields $\mathcal{O}(n_{dof})$.
For common meshes, $\mathcal{O}(n_{el}) \approx \mathcal{O}(n_{dof})$ is a reasonable estimate \cite{graham2006nodal}.
(2) Usually, an iterative solver is employed for solving the resulting sparse linear system of equations.
In the ideal case,
the associated computational complexity is $\mathcal{O}(n_{dof} \, n_{iter})$ \cite{zhou2013linear,greisen2013evaluation}.
Since 
increasing the number of unknowns $n_{dof}$ typically brings along a moderately growing
number of solver iterations $n_{iter}$,
the total complexity of the structural subproblem is expected to be superlinear, but significantly lower than 
quadratic.

\section{Numerical Results} \label{Sec:Results}

\subsection{Elastic Pressure Tube}

\begin{figure}[h!]
	\centering
	\subfloat[Configuration of the elastic tube test case.] 
	{
		\resizebox{0.40\textwidth}{!}{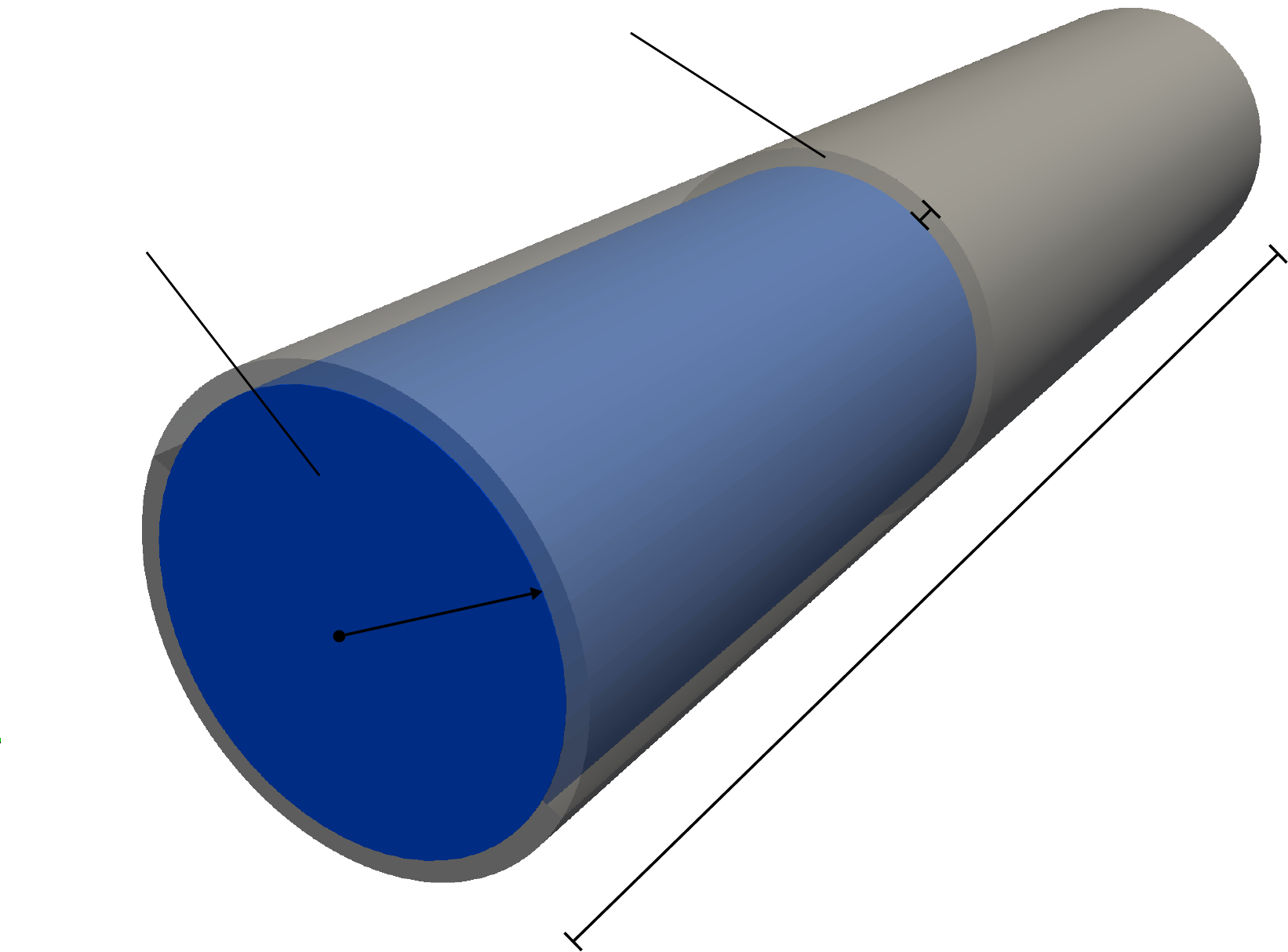 } \label{fig:PressureTubeCase}
	}  
	\quad
	\subfloat[Pressure pulse imposed on the inlet.]
	{
		\resizebox{0.45\textwidth}{!}{\includegraphics[trim=0 20 0 0, clip, width=0.6\textwidth]{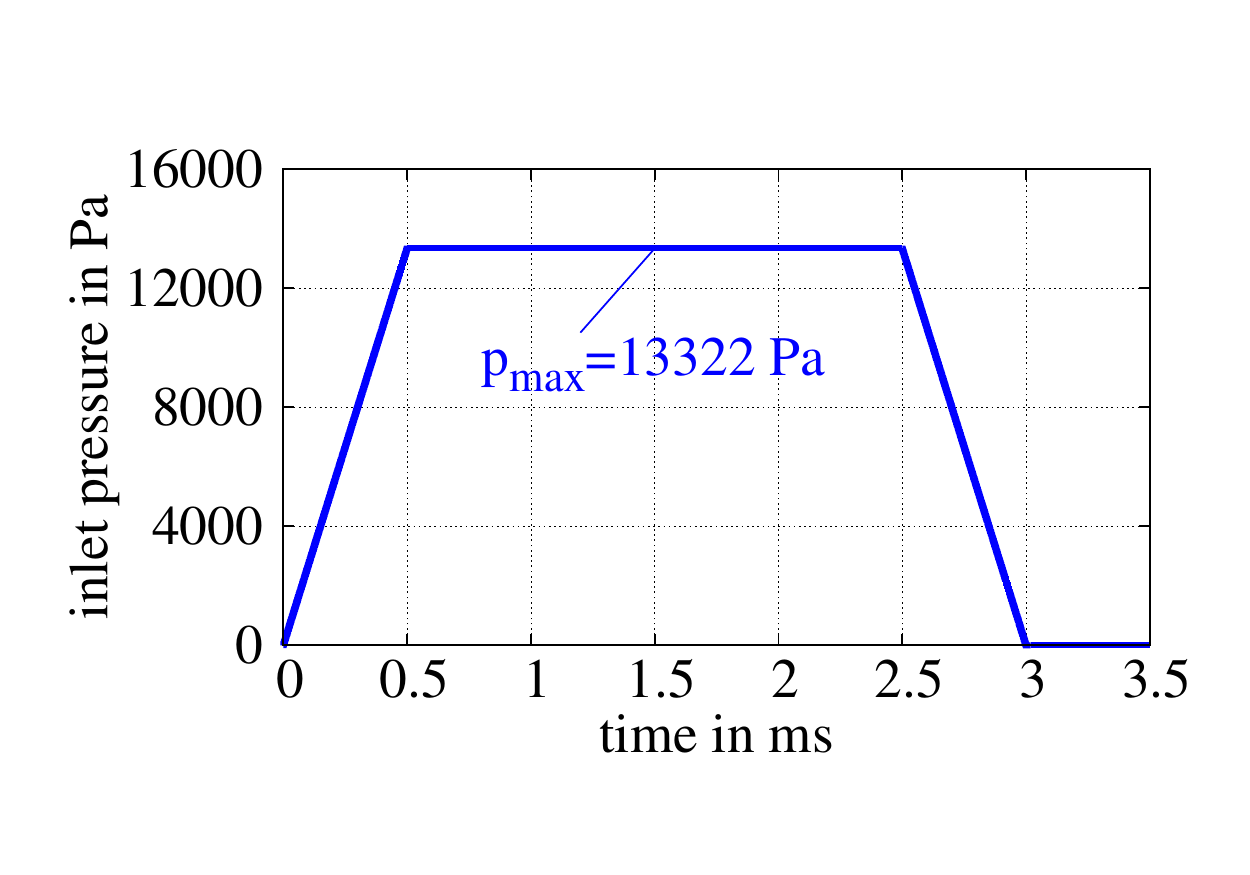} } \label{fig:PressurePulse}
	}  
	\caption{Elastic tube test case}
\end{figure}

The first test case considers an elastic cylindrical tube that is filled with an incompressible fluid,
as depicted in Figure \ref{fig:PressureTubeCase}.
Caused by a short excitation in the beginning,
a pressure pulse propagates through the pipe structure.
While
the configuration is inspired by similar test cases 
discussed, among others, 
by Degroote et al. \cite{degroote2010performance} and Lindner et al. \cite{ComparisonQuasiNewton},
the prescribed pressure pulse is increased by a factor of ten.

The tube has a length of $L = 0.05\,m$, an inner radius of $R=0.005\,m$, and a wall thickness of $s=0.001\,m$.
The elastic structure is characterized by the density
$\rho^s=1200\,\frac{kg}{m^3}$, its Young's modulus $E = 3.0 \cdot 10^5 \frac{kg}{m ~ s^2}$, and a Poisson ratio of $\nu=0.3$;
the fluid's density is $\rho^f=1000\,\frac{kg}{m^3}$ and its dynamic viscosity
$\mu^f=0.001 \frac{kg}{m ~ s}$.
Based on the density ratio of $\rho^f / \rho^s = 0.833$, a strong added-mass instability is to be expected.

Both ends of the channel allow a free in- and outflow of the fluid.
While a constant pressure of $0.0\,Pa$ is prescribed on one end,
the other one is excited by a short pressure pulse with a peak of  $13322.0 \,Pa$
in the beginning of the simulation, see Figure \ref{fig:PressurePulse}.
After that, its boundary pressure is held fixed at $0.0 \, Pa$ as well. 

The fluid domain is discretized by $19193$ tetrahedral elements with $4231$ nodes per time level, 
i.e., a total of $8462$ nodes due to the space-time formulation. The mesh is locally refined in the vicinity of the walls and near the prescribed pressure pulse.
The elastic structure is clamped at both ends; it is represented by $16 \times 40 = 640$ nonlinear isogeometric shell elements defined by a quadratic NURBS.
This spline surface has $714$ control points and $4284$ degrees of freedom.
The simulation runs till $T=8\, ms$ in $80$ time steps of size $\Delta t = 0.1 \,ms$. 
The convergence of the coupling scheme is
detected by a combination of an absolute bound $\varepsilon_{abs}=10^{-8}$
and a relative criterion $\varepsilon_{rel}=10^{-3}$ for the norm 
of the fixed-point residual $\vekt{R}^k$.

\subsubsection{Results}

\begin{figure}[h!]
	\centering
	\subfloat[$t=0.003\,s$: The peak detaches from the inlet.]
	{
				\includegraphics[trim= 0 0 0 0,clip, width=0.65\textwidth]{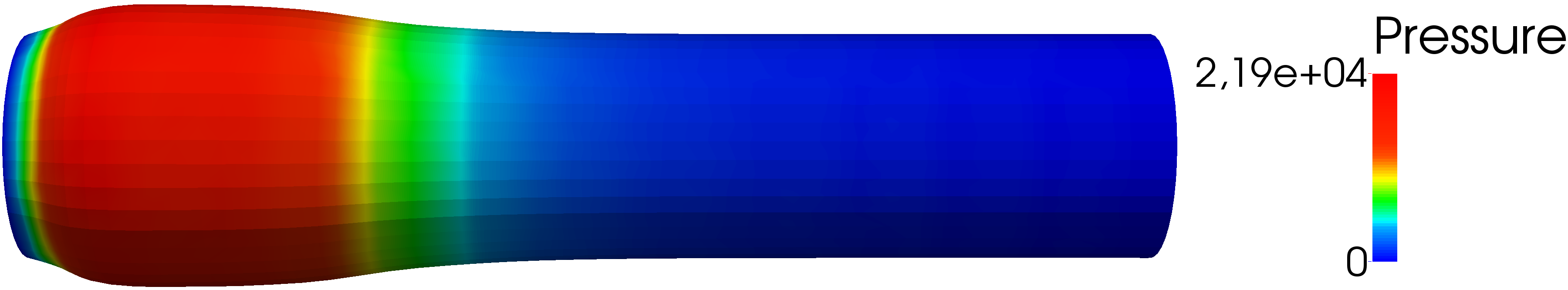}
	}  
	\\
	\subfloat[$t=0.005\,s$: The pressure wave propagates through the structure.]
	{
		\includegraphics[trim=0 -50 0 0,clip, width=0.65\textwidth]{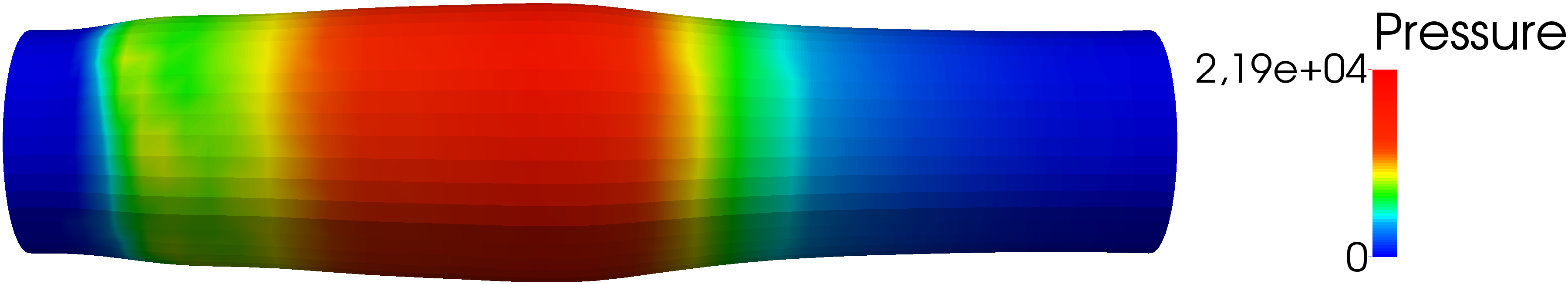}
	}  
	\\
	\subfloat[$t=0.008\,s$: Eventually, it reaches the outflow end of the tube.]
	{
		\includegraphics[trim= 0 -40 0 0,clip, width=0.65\textwidth]{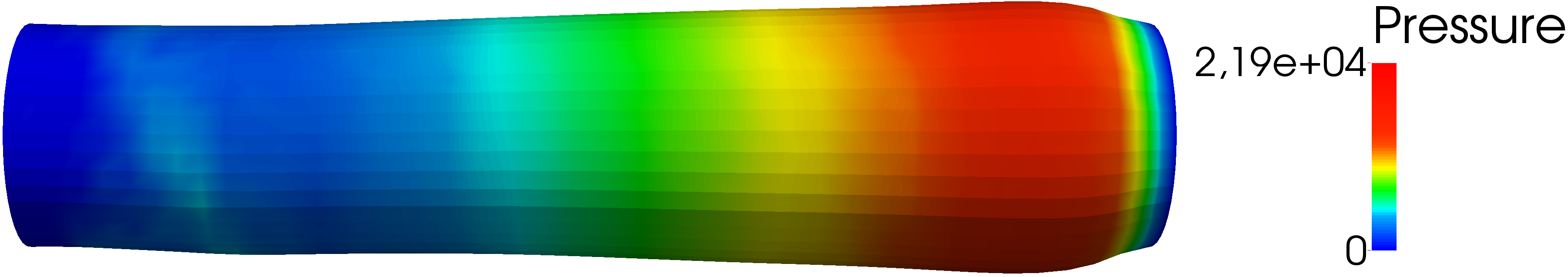}
	}  
	\caption[Elastic Tube: Propagating Shock]{Illustration of the pressure peak running through the tube. Aside from the pressure field, the deformation of the fluid mesh is shown.}
	\label{fig:DeformedTube}
\end{figure}

As stated above, the excitation causes a pressure pulse propagating through the elastic tube,
which is depicted in Figure \ref{fig:DeformedTube} for three sample time levels. 
The snapshots show that 
the moving pressure peak is accompanied by a large widening of the structure.
While retaining its basic profile on its way through the pipe, 
the pulse clearly exhibits some diffusive flattening.
Qualitatively, the numerical results
agree with both physical expectations and
the discussions of similar test cases in literature \cite{Bogaers_IQNMVJ,PerformancePartitionedMonolithic,ComparisonQuasiNewton}.
Note that 
this solution is independent (within the chosen convergence tolerance) from the update technique employed.

The focus of this test case
is on the efficiency of different updating schemes.
A comparison is provided in Table \ref{Tab:PressureTubeComparison1}:
Aside from the average number of coupling iterations per time step, 
it lists
the relative speed-up in terms of coupling iterations and runtime
with respect to a constant under-relaxation;
lastly, the percentages of computation time spent for the interface quasi-Newton update are 
indicated.
\begin{table}[h!]
	\centering
	\begin{tabular}{ | l | c | c | c |  c |}
		\hline
		\multirow{2}{*} {Method} 	& Average Coupling & Relative Coupling  & Relative & IQN-Time /   \\
					  &  Iterations  &   Iterations & Runtime & Runtime\\
		\hline \hline
		Under-relaxation, $\omega=0.05$ & 100.39	& 100.00\,\%& 100.00\,\% & -\\
		Aitken's relaxation				& ~\,26.38		& ~\,26.28	\% 		& ~\,26.31\,\% 		& - \\
		\hline 
		IQN-ILS ($q=0$) & ~\,14.96 		& ~\,14.90\,\% 		& ~\,14.93\,\% 		 &  ~\,0.087\,\% \\
		IQN-ILS, $q=1$ 	& ~\,~\,8.91  	 & ~\,~\,8.88\,\% 	& ~\,~\,9.18\,\% 	 & ~\,0.122\,\% \\
		IQN-ILS, $q=5$  	& ~\,~\,8.06 	& ~\,~\,8.03\,\% 	& ~\,~\,8.33\,\% 	& ~\,0.426\,\% \\
		IQN-ILS, $q=10$ & ~\,~\,8.66 	& ~\,~\,8.63\,\% 	& ~\,~\,9.03\,\% 	& ~\,1.042\,\% \\
		IQN-ILS, $q=20$ & ~\,~\,9.76		& ~\,~\,9.72\,\% 	& ~\,10.14\,\% 		& ~\,2.890\,\% \\
		\hline 
		IQN-IMVJ 			& ~\,~\,8.11 		& ~\,~\,8.08\,\% 	& ~\,~\,9.53\,\% 	& 11.736\,\%\\
		\hline
		IQN-MVJ-RS-SVD, $M=5$, $\varepsilon_{svd}=0.01$   & ~\,~\,8.09		& ~\,~\,8.06\,\% 	& ~\,~\,8.41\,\% 	& ~\,0.370\,\%\\
		\hline
		IQN-IMVLS, explicit Jacobian 	& ~\,~\,8.14 	& ~\,~\,8.10\,\% 	& ~\,~\,8.50\,\% 	& ~\,1.263\,\% \\
		IQN-IMVLS, $q=5$ 	& ~\,~\,8.54 	& ~\,~\,8.50\,\% 	&  ~\,~\,8.93\,\% 	& ~\,0.070\,\% \\
		IQN-IMVLS, $q=10$ 	& ~\,~\,8.28 	& ~\,~\,8.24\,\%	&  ~\,~\,8.68\,\% 	& ~\,0.078\,\% \\
		IQN-IMVLS, $q=20$ 	& ~\,~\,7.95 	& ~\,~\,7.92\,\% 	&  ~\,~\,8.28\,\%	& ~\,0.089\,\% \\
		IQN-IMVLS, $q=50$ 	& ~\,~\,8.05 	& ~\,~\,8.02\,\%	& ~\,~\,8.34\,\% 	& ~\,0.113\,\% \\
		IQN-IMVLS, $q=80$ 	& ~\,~\,8.05 	& ~\,~\,8.02\,\%	& ~\,~\,8.35\,\% 	& ~\,0.120\,\% \\
		\hline
		IQN-IMVLS, $q=80$, 1 explicit step &  ~\,~\,7.76 &  ~\,~\,7.73\,\% & ~\,~\,8.02\,\% & ~\,0.253\,\%  \\
		\hline
	\end{tabular}
	\caption[] {Comparison of different update schemes for the pressure tube test case.}
	\label{Tab:PressureTubeComparison1}
\end{table}
The first observation is that,
although Aitken's dynamic relaxation converges much faster
than the constant under-relaxation,
neither of them is really
feasible for this test case, since they require about $100$ or $26$ coupling iterations
per time step,
respectively.
The problem is the high added-mass instability,
requiring very small relaxation factors;
in this case, $\omega \approx 0.05$ was found to be the highest value
precluding divergence.

As opposed to this,
all employed interface quasi-Newton variants 
entail a substantial 
increase in performance
-- both in terms of reducing the number of coupling iterations and the runtime.
In particular, the incorporation of data from past time steps proves very beneficial.
With this, using an IQN approach reduces the 
total runtime compared to a constant relaxation by up to $92 \%$;
perhaps even more remarkable is the 
decrease by about $70 \%$ with respect to Aitken's method.

The results for the IQN-ILS variant
identify its dependency on the number of explicitly reused time steps $q$:
Away from the optimum of this test case,
$q \approx 5$,  
including either too few or too many previous steps compromises the effectiveness;
in fact, the coupling scheme was even diverging for values of $q$ higher than about $25$.
Note in this context that no filtering technique is applied within this work.

Against this backdrop, the implicit incorporation of past data
via the previous inverse Jacobian 
deserves special notice:
Despite not requiring any problem-dependent parameters,
it
enables the IQN-IMVJ variant to
keep up with the optimum of the IQN-ILS approach in terms of coupling iterations.
However, the actual runtime indicates its major drawback:
While the computational cost of the IQN-ILS method barely exceeds $1\,\%$ of the total runtime
for moderate choices of $q$,
handling the explicit Jacobian approximation raises the effort for the IQN-IMVJ method
to $11.74\,\%$; hence, its cost is by no means negligible against the solver calls.

That is exactly what the IQN-IMVLS variant is designed to provide a remedy for.
Since the analytic form of the quasi-Newton update stays untouched,
the required coupling iterations 
of the IQN-IMVJ and the IQN-IMVLS approach with an explicit Jacobian
show only an insignificant difference arising from numerical rounding.
The IQN-IMVLS method,
however,
employs more advantageous numerical techniques and 
a beneficial storage of already determined quantities
(as discussed in Section \ref{SubSec:IQN-IMVLS})
to
reduce the update's computational cost.
In particular, the number of operations involving an explicit Jacobian approximation
is reduced.
The results in Table \ref{Tab:PressureTubeComparison1}
prove that these adjustments entail a significant 
speed-up of the multi-vector IQN approach  --
in this case by a factor of ten.

Moreover, 
the IQN-IMVLS method also provides 
a purely implicit version avoiding any explicit 
representation of the Jacobian approximation.
This variant reintroduces 
the number of past time steps $q$ to be incorporated;
in contrast to the explicit reutilization, however, 
this parameter is not very problem-dependent or critical,
as high values of $q$ do not entail any numerical problems.
Instead, the quality of the update scheme in general benefits from an
increasing $q$,
as confirmed by Table \ref{Tab:PressureTubeComparison1}:
Although there
seems to be a marginal optimum at around $q=20$ 
for this test case, 
the convergence speed  afterwards
aligns more and more
with the one observed for 
including all past time steps ($q=80$).
In fact, 
it remains virtually unchanged for values higher than $q=50$.
In combination with the computational effort, which stays at around 
$0.1\,\%$ of the runtime even for including all previous steps,
using the purely implicit update further improves the IQN-IMVLS method
for this setting. 
Without compromising effectiveness,
the restart-based 
IQN-MVJ-RS-SVD method significantly reduces the cost of the multi-vector approach, too.
Although being slightly more expensive than the new algorithm for this test case, 
the observed difference is small enough to be blurred by implementation details and parameter choices to some extent;
in relation to the overall simulation time, both methods come at negligible cost.
When it comes to including the most recent time step explicitly in the
IQN-IMVLS method as suggested in Remark \ref{Remark2}, the last row in Table \ref{Tab:PressureTubeComparison1}
confirms that this option does in fact slightly reduce
the number of coupling iterations without being too expensive. The effect, however,
is not very marked. \\

As indicated before,
the main advantage of the purely implicit update,
i.e., its linear complexity,
becomes more distinct
for larger problem scales.
In general, the cost of the discussed interface quasi-Newton concept
depends on the number of structural degrees of freedom
at the FSI interface.
As the elastic tube is modeled by shell elements, however,
in this test case the FSI interface is equivalent to the whole structural domain.
For shell analysis,
a distinction between all structural degrees of freedom
and those at the interface is redundant.

\begin{figure}[h!]
	\centering
	\includegraphics[trim=0 0 0 10, clip, width=0.8\textwidth]{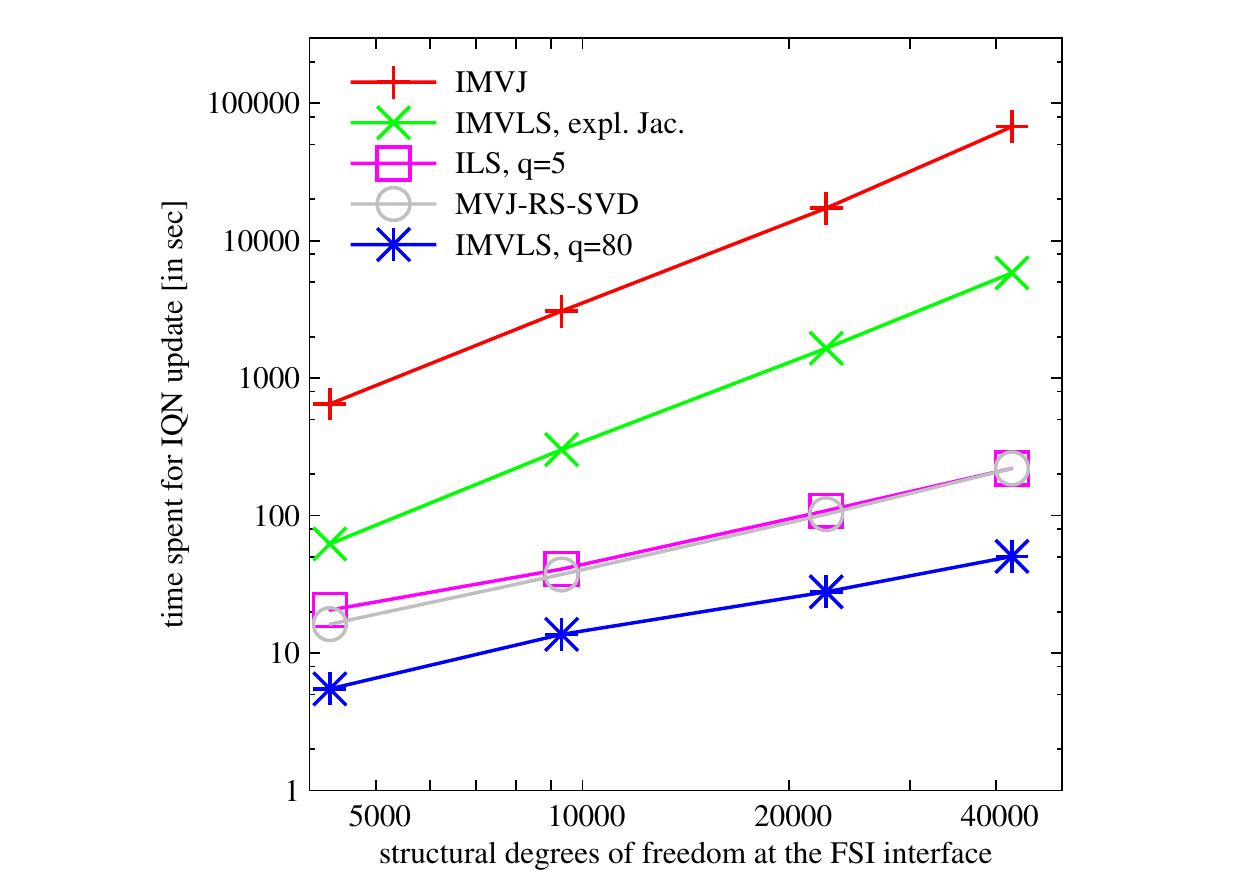}   \label{Fig:PlotA}
	\caption[]{Double-logarithmic plot of the computational cost for different interface quasi-Newton variants in dependence on the interface resolution.}
	\label{Fig:Plot}
\end{figure}

Therefore, 
Figure \ref{Fig:Plot} plots the absolute time spent for the multi-vector IQN variants
as well as the IQN-ILS method with $q=5$ 
over the 
number of structural degrees of freedom.
To assess the complexity of the different methods, both axes are scaled logarithmically.
The first observation is
the exploding cost of the IQN-IMVJ method, caused
by its extensive usage of the explicit Jacobian approximation.
While
the IQN-IMVLS method with an explicit inverse Jacobian again
significantly reduces the numerical effort,
it does not keep the cost from growing quadratically with
the problem scale.
In contrast, 
the purely implicit IQN-IMVLS version is not only much cheaper;
the plot clearly points out its linear scaling with the interface resolution.
Moreover, for this test setting it outperforms both the IQN-MVJ-RS-SVD and the IQN-ILS method, 
which also show a linear complexity.

\begin{table}[h!]
	\centering
	\begin{tabular}{ | l | c | c |}
		\hline 
		\multirow{2}{*} {Method} 	& Average Coupling & IQN / Structure  \\
		&  Iterations   & in\,\% \\
		\hline \hline
		\multicolumn{3}{| c |}{$9324$ Structural Degrees of Freedom (at the FSI Interface)} \\
		\hline \hline
		IQN-ILS, $q=5$ & 7.84 &  ~\,0.37\,\% \\
		IQN-IMVJ & 8.01  &  27.36\,\%\\
		IQN-MVJ-RS-SVD, $M=5$, $\varepsilon_{svd}=0.01$  & 7.99 &  ~\,0.35\,\%\\
		IQN-IMVLS, explicit Jacobian & 8.05& ~\,2.66\,\%\\
		IQN-IMVLS, $q=80$ & 7.99 &   ~\,0.12\,\% \\
		IQN-IMVLS, $q=80$, 1 explicit step&  7.48 &   ~\,0.22\,\%\\
		\hline \hline
		\multicolumn{3}{| c |}{$22644$ Structural Degrees of Freedom (at the FSI Interface)} \\
		\hline \hline
		IQN-ILS, $q=5$ & 8.20 &  ~\,0.33\,\%\\
		IQN-IMVJ & 7.91 &  53.25\,\% \\
		IQN-MVJ-RS-SVD, $M=5$, $\varepsilon_{svd}=0.01$  & 8.09 &  ~\,0.32\,\%\\
		IQN-IMVLS, explicit Jacobian & 7.99  & ~\,5.08\,\%\\
		IQN-IMVLS, $q=80$ & 7.86  &  ~\,0.09\,\%\\
		IQN-IMVLS, $q=80$, 1 explicit step & 7.79 &  ~\,0.19\,\% \\
		\hline \hline
		\multicolumn{3}{| c |}{$42228$ Structural Degrees of Freedom (at the FSI Interface)} \\
		\hline \hline
		IQN-ILS, $q=5$ & 8.21 &  ~\,0.27\,\%\\
		IQN-IMVJ & 7.99 &  83.58\,\% \\
		IQN-MVJ-RS-SVD, $M=5$, $\varepsilon_{svd}=0.01$  &  8.04 &  ~\,0.27\,\%\\
		IQN-IMVLS, explicit Jacobian & 7.88 &   ~\,7.31\,\%\\
		IQN-IMVLS, $q=80$ & 7.97 & ~\,0.06\,\%\\
		IQN-IMVLS, $q=80$, 1 explicit step & 8.20 &  ~\,0.15\,\%\\
		\hline
	\end{tabular} 
	\caption[] {Comparison of IQN variants for three refinement levels. Due to the shell model, all structural degrees of freedom are at the FSI interface. }
	\label{Tab:PressureTubeComparison2}
\end{table}

While Figure \ref{Fig:Plot} compares the efficiency of different quasi-Newton schemes,
it does not allow any judgment on to which extent their computational cost is negligible compared
to the solver calls.
Therefore,
Table \ref{Tab:PressureTubeComparison2} 
puts them into context:
Taking into account that the fluid mesh does not have a direct influence on the IQN approach
and that the structural domain is equivalent to the FSI interface due to the usage of shell elements,
its last column relates the time spent for the quasi-Newton methods
to that required for the  structural solution; 
it considers three different levels of refinement of the structural discretization,
i.e., $9234$, $22644$, and $44228$ degrees of freedom.
Moreover, Table \ref{Tab:PressureTubeComparison2} lists the average numbers of coupling iterations per time step,
which stay close to $8$ for all tests conducted.
Nevertheless,
the investigated methods show vast differences regarding their computational cost.

Again, the most striking observation is the very high cost of the IQN-IMVJ method:
For the three refinement levels, 
the runtime it accounts for ranges from $27.35 \%$ to $83.58 \%$  compared to the time required for the structural solution,
and therefore seriously slows down the whole simulation.
Moreover, this percentage is growing with the problem scale, indicating that the quadratic complexity causes the cost of the IQN-IMVJ approach to be increasing faster than that
of solving the structural subproblem.
As discussed before, the IQN-IMVLS method with an explicit Jacobian approximation significantly
reduces the numerical effort, but not the method's complexity;
hence, its relative cost is still growing with the number of degrees of freedom, in this case from $2.66\%$ to $7.31\%$.
In contrast, 
the complexity of the purely implicit IQN-IMVLS version,
the IQN-MVJ-RS-SVD method, and the IQN-ILS approach is linear.
As a consequence, their portion on the total runtime is not only negligible for all the conducted
simulations,
but even decreasing with the structural refinement.
In accordance to the discussion in Remark \ref{Remark3}, these observations confirm that the complexity of solving the structural subproblem
is between being linear and quadratic for this setting.

For the first and the second refinement levels,
the explicit incorporation of the most recent time step in the least-squares problem
of the IQN-IMVLS method again
slightly accelerates the coupling.
Unfortunately, the lower convergence speed for the finest level
stands in the way of a general recommendation.

All in all, the purely implicit IQN-IMVLS variant
proves to overcome the main drawback of the IQN-IMVJ method,
i.e., the quadratic complexity,  
without making compromises regarding the convergence speed or
the lack of parameter-tuning.

\subsection{Sloshing Tank}

The second example is a
cylindrical deformable tank partially filled with an incompressible fluid. 
While Figure \ref{fig:SloshinTank}  illustrates the basic configuration, 
the test case parameters are listed in Table \ref{tab:SloshingTank}. 

The simulation is started from the steady state where the fluid is at rest
and the deformation has adjusted to the hydrostatic pressure.
During the simulation, the system is excited in horizontal direction by a periodic movement of the tank bottom 
with the prescribed displacement $\Delta x(t) = A \, \sin (2 \pi f \,t)$.
Due to inertia,
this excitation results in complex
and highly intertwined motions of the deforming tank structure and the liquid sloshing inside.

The evolving fluid domain is discretized by a space-time mesh of $16243$ elements 
and $4024$ nodes per time level;
since it is adapted to the free-surface via interface tracking \cite{NorbertPhD,Steffi2015},
the tank wall and its bottom involve slip-conditions
for both the fluid and the mesh.
The cylindrical tank wall is modeled by $32 \times 30 =960$ nonlinear isogeometric shell elements of degree $2$ defined by a NURBS with $1184$ control points. 
The simulation runs for $1600$ time steps of size $\Delta t=0.01$s;
the absolute and relative convergence criteria for the interface deformation are $\varepsilon_{abs}=10^{-5}$ and $\varepsilon_{rel}=10^{-4}$.

\begin{minipage}[l]{0.42\textwidth}
	\vspace{0.85cm}
	\resizebox{0.79\textwidth}{!}{
\begingroup%
  \makeatletter%
  \providecommand\color[2][]{%
    \errmessage{(Inkscape) Color is used for the text in Inkscape, but the package 'color.sty' is not loaded}%
    \renewcommand\color[2][]{}%
  }%
  \providecommand\transparent[1]{%
    \errmessage{(Inkscape) Transparency is used (non-zero) for the text in Inkscape, but the package 'transparent.sty' is not loaded}%
    \renewcommand\transparent[1]{}%
  }%
  \providecommand\rotatebox[2]{#2}%
  \ifx\svgwidth\undefined%
    \setlength{\unitlength}{160.0bp}%
    \ifx\svgscale\undefined%
      \relax%
    \else%
      \setlength{\unitlength}{\unitlength * \real{\svgscale}}%
    \fi%
  \else%
    \setlength{\unitlength}{\svgwidth}%
  \fi%
  \global\let\svgwidth\undefined%
  \global\let\svgscale\undefined%
  \makeatother%
  \begin{picture}(1,0.85693664)%
    \put(0,0){\includegraphics[width=\unitlength,trim=0 0 0 0]{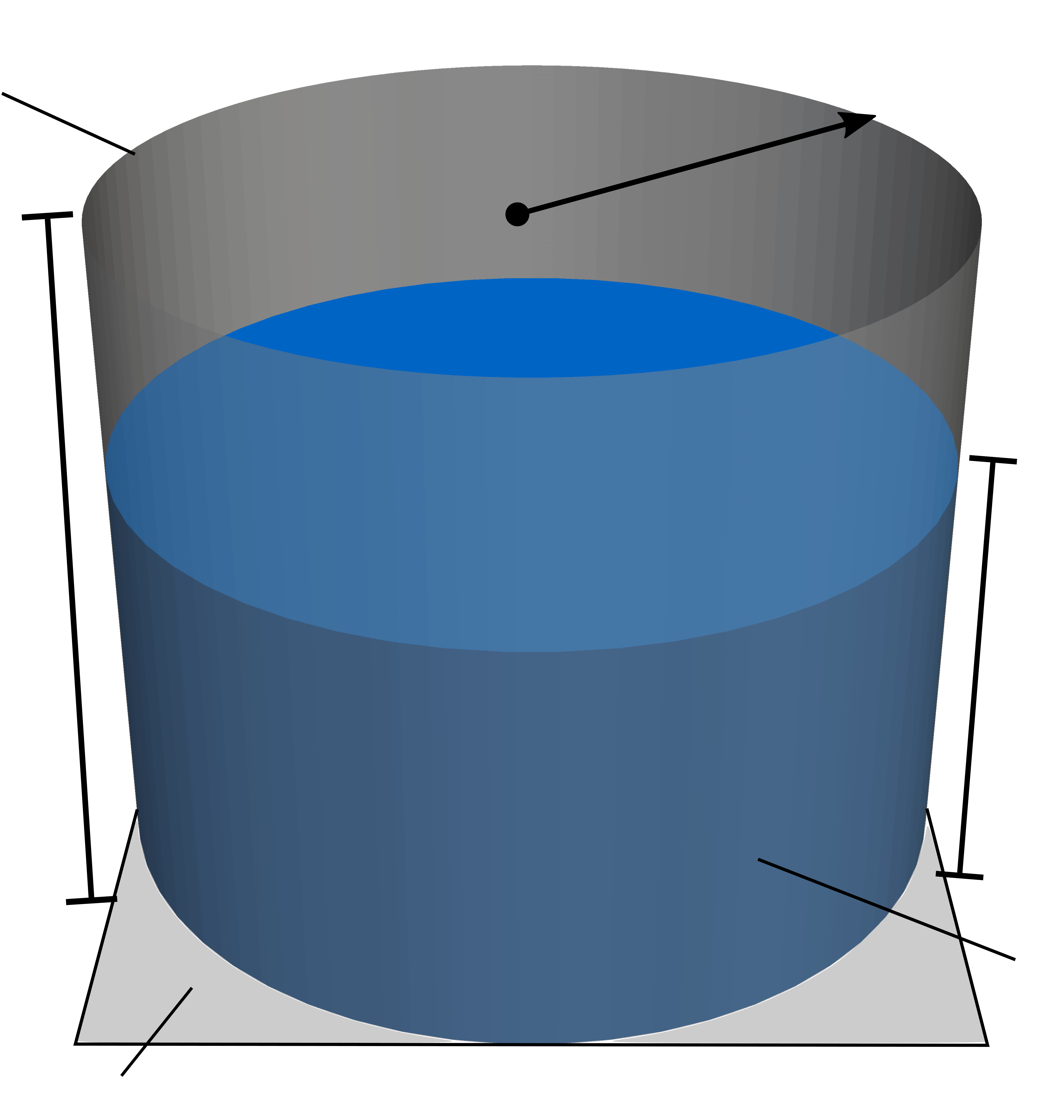}}%
    \put(0.95,0.42){\color[rgb]{0,0,0}\makebox(0,0)[lt]{\begin{minipage}{0.19456515\unitlength}\raggedright $h$ \end{minipage}}}%
    \put(0.97,0.10){\color[rgb]{0,0,0}\makebox(0,0)[lb]{\smash{Fluid}}}%
    \put(-0.05,0.54204183){\color[rgb]{0,0,0}\makebox(0,0)[lt]{\begin{minipage}{0.19456515\unitlength}\raggedright $H$ \end{minipage}}}%
    \put(0.65,0.88){\color[rgb]{0,0,0}\makebox(0,0)[lt]{\begin{minipage}{0.19456515\unitlength}\raggedright $R$\end{minipage}}}%
    \put(-0.1,0.98){\color[rgb]{0,0,0}\makebox(0,0)[lb]{\smash{Elastic Wall}}}%
    \put(0.0,-0.03){\color[rgb]{0,0,0}\makebox(0,0)[lb]{\text{Moving Base}}}%
 \end{picture}%
\endgroup%

	}
	\captionof{figure}{Sloshing tank test case.}
	\label{fig:SloshinTank}
\end{minipage}
\begin{minipage}{0.45\textwidth}
	\centering
	\begin{tabular}{ | l | l | c | c |}
		\hline 
		\textbf{Field} & \textbf{Parameter} & \textbf{Symbol} & \textbf{Value} \\
		\hline \hline 
		Fluid & Density & $\rho^f$ &  $900\,\frac{kg}{m^3}$ \\
		\parbox[][0.5cm]{0pt}	& Dynamic viscosity  & $\mu^f$ &  $0.4\,\frac{kg}{m \, s}$ \\
		\hline
		Structure & Density & $\rho^s$ & $7856\,\frac{kg}{m^3}$\\
		& Young's modulus &    $E$ &  $2.15\text{E+9}\,\frac{kg}{m \, s^2}$ \\
		& Poisson ratio &    $\nu$ &  $0.3$ \\
		\hline
		Geometry & Radius & $R$ &  $50\,m$ \\
		& Tank height & $H$ & $80\,m$ \\
		& Filling level & $h$ & $50\,m$ \\
		& Wall thickness & $s$ & $0.01\,m$ \\
		\hline
		Excitation & Amplitude & $A$ & $5.0\,m$ \\
		& Frequency & $f$ & $0.0625 \,Hz$ \\
		\hline
	\end{tabular}
	\captionof{table}{Test case parameters.}
	\label{tab:SloshingTank}  
\end{minipage}  

\subsubsection{Results}

\begin{figure}[h!]
	\centering
	\subfloat[$t=6.0\,s$] 
	{
		\includegraphics[trim=0 0 0 0, clip, width=0.28\textwidth]{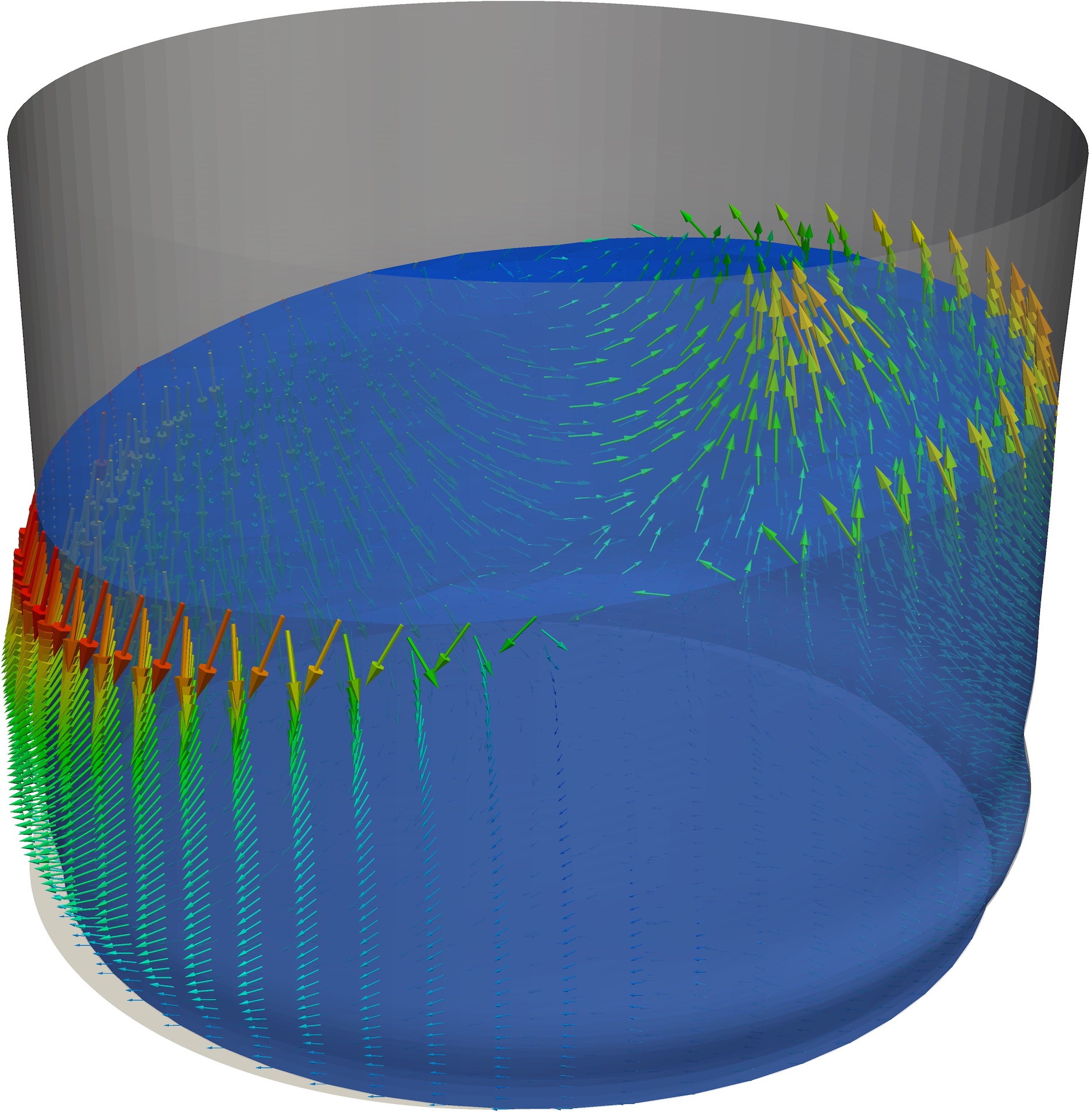} 
	}  
	\quad
	\subfloat[$t=10.0\,s$]
	{
			\includegraphics[trim=0 0 0 0,  clip, width=0.28\textwidth]{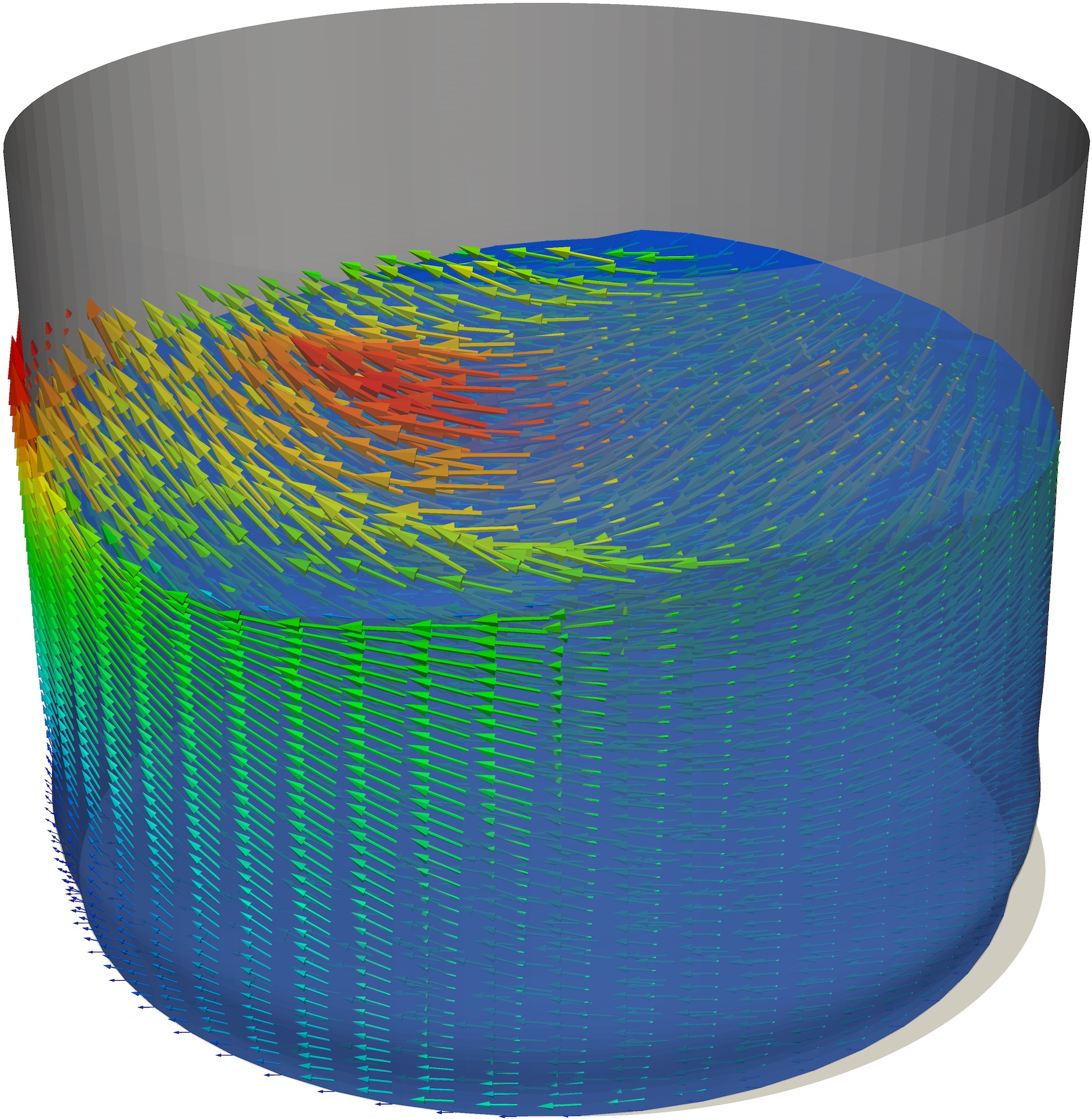} 
	}  
	\quad  
	\subfloat[$t=15.0\,s$]
	{
		\includegraphics[trim=0 0 0 0,  clip, width=0.28\textwidth]{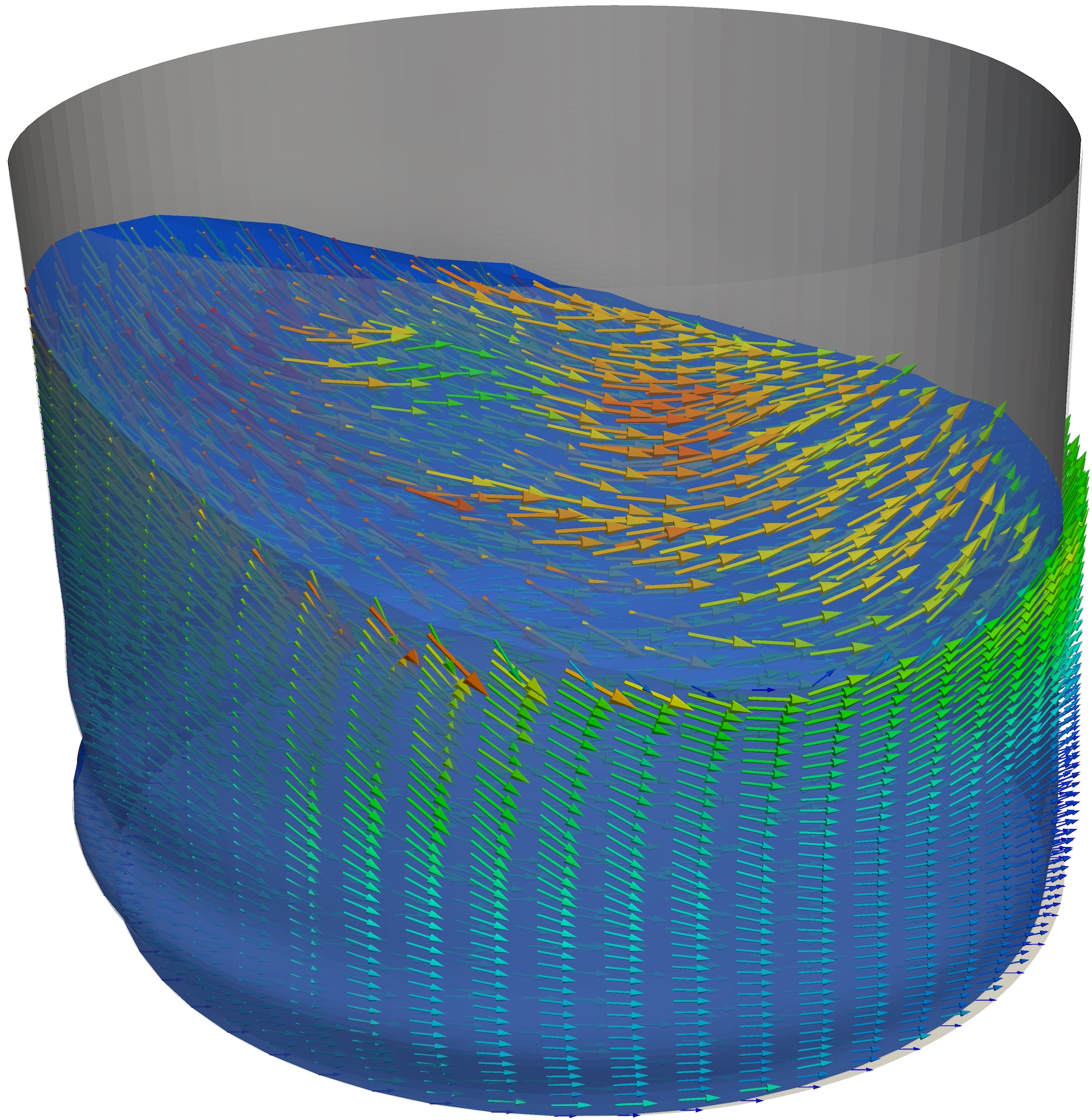} 
	}  
	\caption[]{Snapshots of the sloshing tank test case. The colored arrows visualize the velocity field.}
	\label{fig:SnapshotsTank}
\end{figure}

To give an impression of the simulation results, Figure \ref{fig:SnapshotsTank} depicts
three different snapshots of the sloshing tank. Both the free surface and the tank structure exhibit large displacements. 

\begin{table}[h!]
	\centering
	\begin{tabular}{ | l | c |  c |}
		\hline
		Method 	& Average Coupling Iterations & IQN-Time /  Runtime  \\
		\hline \hline
		Under-relaxation, $\omega=0.1$ &  81.64	&  -\\
		Aitken's relaxation& 26.16 & 	 - \\
		\hline 
		IQN-ILS ($q=0$) & 17.90 &  ~\,0.08\,\% \\
		IQN-ILS, $q=1$ &   ~\,8.65&    ~\,0.08\,\% \\
		IQN-ILS, $q=5$ &  ~\,4.97 &    ~\,0.19\,\% \\
		IQN-ILS, $q=10$ &  ~\,4.45 &    ~\,0.39\,\% \\
		IQN-ILS, $q=20$ &  ~\,4.17 &  ~\,0.64\,\% \\
		IQN-ILS, $q=50$ &   ~\,4.20 &  ~\,2.29\,\% \\
		IQN-ILS, $q=100$ &   ~\,4.40 &   ~\,8.58\,\% \\
		\hline 
		IQN-IMVJ & ~\,5.10  &   11.33\,\%\\
		\hline
		IQN-MVJ-RS-SVD, $M=5$, $\varepsilon_{svd}=0.01$ & ~\,5.18 &  ~\,1.28\,\%\\
		\hline
		IQN-IMVLS, explicit Jacobian & ~\,5.12  &  ~\,1.67\,\% \\
		IQN-IMVLS, $q=50$ &  ~\,7.14 &  ~\,0.08\,\% \\
		IQN-IMVLS, $q=100$ & ~\,6.23  &   ~\,0.12\,\% \\
		IQN-IMVLS, $q=200$ & ~\,5.56   &   ~\,0.20\,\% \\
		IQN-IMVLS, $q=500$ &  ~\,5.05  &   ~\,0.37\,\% \\
		IQN-IMVLS, $q=1000$ & ~\,5.02 & ~\,0.49\,\% \\
		IQN-IMVLS, $q=1600$ & ~\,5.11  &    ~\,0.67\,\% \\
		\hline
		IQN-IMVLS, $q=500$, 1 explicit step & ~\,4.05  &  ~\,0.62\,\% \\
		IQN-IMVLS, $q=1600$, 1 explicit step & ~\,4.21  &  ~\,1.03\,\% \\
		\hline
	\end{tabular}
	\caption[] {Comparison of different update techniques for the sloshing tank test case.}
	\label{Tab:TankComparison}
\end{table}
The efficiency of different update techniques is compared in Table \ref{Tab:TankComparison},
both in terms of coupling iterations and their computational cost.
All in all, the results are in very good agreement with those of the elastic tube scenario:
While neither a constant nor Aitken's dynamic relaxation converge fast enough to be a reasonable option,
all interface quasi-Newton variants drastically speed up the coupling scheme.
Again, incorporating information from past time steps is extremely beneficial.
The IQN-ILS method performs very well for this test case, showing a quite flat optimum of choosing the parameter $q$;
the cost of solving the least-squares problem, however, is clearly increasing with $q$.

Although being slightly less efficient for this test case, 
the
implicit reutilization of the multi-vector methods almost keeps up with the IQN-ILS approach in terms of the required coupling iterations.
While the IQN-IMVJ variant pays for this fast convergence with the expensive explicit Jacobian, the computational cost of 
both the IQN-IMVLS and the IQN-MVJ-RS-SVD method
is comparable to that of the IQN-ILS method, thanks to the linear complexity.
In direct comparison, 
the new IQN-IMVLS method outruns the restart-based variant for this test case;
however, this statement is again best put into the perspective of the specific implementation and the chosen parameters.

Increasing the value $q$ brings the convergence speed closer to that of the explicit multi-vector approach.
Although a rather high value is required here to reach this speed,
the implicit Jacobian product proves to be very cheap as 
the computational cost stays non-critical even for the maximum of reusing all time steps, i.e., $q=1600$.
By including the data pairs of the most recent time step explicitly in the least-squares problem,
for this test case
the convergence of the IQN-IMVLS method is accelerated by about one coupling iteration 
per time step
at moderate cost.

Altogether, the test case proves that the IQN-IMVLS method
competes with the IQN-ILS approach
not only in terms of the convergence speed, but also regarding the computational cost.

\section{Conclusion}

This paper presents a novel interface quasi-Newton method improving 
the partitioned simulation of
fluid-structure interaction.
Building on the inverse multi-vector Jacobian (IQN-IMVJ) variant, 
it
overcomes the main drawback: As the cost of handling an explicit 
inverse Jacobian approximation increases quadratically with the problem size,
the IQN-IMVJ method significantly slows down large-scale simulations.
The new interface quasi-Newton
implicit multi-vector least-squares
(IQN-IMVLS) method, in contrast, 
systematically avoids any explicit Jacobian approximation,
so that
the multi-vector update is realized with linear complexity.
As a consequence, its computational cost stays negligible
even and particularly for industrial-scale applications.
The effectiveness of the IQN-IMVLS method is confirmed by the numerical test cases in Section \ref{Sec:Results}.
Although the number of required coupling iterations stays virtually unchanged
compared to the IQN-IMVJ variant,
the results show a substantial reduction of computational cost.
In accordance to the discussion above, this difference becomes more and more distinct for finer resolutions of the
structural discretization.
One major strength of the 
multi-vector concept is that
its implicit incorporation of past data 
does not rely on problem-dependent parameters -- and yet
keeps up with 
an optimal parameter choice of the 
interface quasi-Newton implicit least-squares (IQN-ILS) method,
regarding the required coupling iterations.
While the IQN-IMVJ approach had to pay a high price 
in computational cost
for this 
capability,
the IQN-IMVLS method 
is comparable to the IQN-ILS variant concerning the runtime. 

Although the IQN-IMVLS variant reintroduces the number of reused time steps $q$,
this parameter is not problematic at all as the quality of 
the quasi-Newton update in general benefits from including more steps.\\

In conclusion, 
the new IQN-IMVLS method combines the advantages
of the IQN-IMVJ and the IQN-ILS variant
without their individual drawbacks.
Coping without critical parameters,
it succeeds in realizing
the favorable implicit reutilization of past data with a computational complexity
that grows linearly with the problem size. 

\appendix

 \section{Proof of New Jacobian Update} 
\label{AppendixProofA}
Via mathematical induction,
this section proves 
the equality of the two Jacobian updates discussed in Section \ref{SubSec:IQN-IMVLS}:
\begin{align}
\JacN{n+1} =
\JacN{n} + \left( \vekt{W}_k^n - \JacN{n} \vekt{V}_k^n \right) \vekt{Z}_k^n  \quad \text{and} \quad
 \JacN{n+1}  = \JacN{0} \prod_{i=0}^n \left( \vekt{I} - \vekt{V}_k^i \vekt{Z}_k^i \right) + \sum_{i=0}^n \vekt{W}_k^i \vekt{Z}_k^i  \prod_{j=i+1}^n \left( \vekt{I} - \vekt{V}_k^j \vekt{Z}_k^j \right) ~.  \label{Eqn:AppendixA1} \nonumber
\end{align}

\begin{enumerate}
	\item Induction base case:	For $n=0$ the relation holds, since the second expression yields
		\begin{align}
	\JacN{1} = 
	 \JacN{0} \prod_{i=0}^0 \left( \vekt{I} - \vekt{V}_k^i \vekt{Z}_k^i \right) + \sum_{i=0}^0 \vekt{W}_k^i \vekt{Z}_k^i \underbrace{\prod_{j=i+1}^0 \left( \vekt{I} - \vekt{V}_k^j \vekt{Z}_k^j \right)}_{ \prod_{p}^{q} = 1 \text{ for } p>q }
	= \JacN{0} \left( \vekt{I} - \vekt{V}_k^0 \vekt{Z}_k^0 \right) + \vekt{W}_k^0 \vekt{Z}_k^0
	= \JacN{0} + \left( \Wk^0 - \JacN{0} \Vk^0 \right) \vekt{Z}_k^0	 ~. \nonumber
	\end{align}
	\item Inductive step: If the equality is satisfied for any $n$,
	it can be shown to hold for $n+1$ via
	\begin{align}
		\JacN{n+1} &= \JacN{n} + \left( \vekt{W}_k^n - \JacN{n} \vekt{V}_k^n \right) \vekt{Z}_k^n = \JacN{n} \left( \vekt{I} - \Vk^n \, \vekt{Z}_k^n  \right) + \Wk^n \, \vekt{Z}_k^n \nonumber \\ \nonumber
		&= \left[   \JacN{0}  \prod_{i=0}^{n-1} \left( \vekt{I} - \vekt{V}_k^i \vekt{Z}_k^i \right) + \sum_{i=0}^{n-1} \vekt{W}_k^i \vekt{Z}_k^i  \prod_{j=i+1}^{n-1} \left( \vekt{I} - \vekt{V}_k^j \vekt{Z}_k^j \right)	\right]		
		\left( \vekt{I} - \Vk^n \, \vekt{Z}_k^n  \right)  + \Wk^n \, \vekt{Z}_k^n  \\ \nonumber
	&=    \JacN{0}  \prod_{i=0}^{n} \left( \vekt{I} - \vekt{V}_k^i \vekt{Z}_k^i \right) + \sum_{i=0}^{n-1} \vekt{W}_k^i \vekt{Z}_k^i  \prod_{j=i+1}^{n} \left( \vekt{I} - \vekt{V}_k^j \vekt{Z}_k^j \right)		
	+ \Wk^n \, \vekt{Z}_k^n  \\ \nonumber
		&=    \JacN{0}  \prod_{i=0}^{n} \left( \vekt{I} - \vekt{V}_k^i \vekt{Z}_k^i \right) + \sum_{i=0}^{n-1} \vekt{W}_k^i \vekt{Z}_k^i  \prod_{j=i+1}^{n} \left( \vekt{I} - \vekt{V}_k^j \vekt{Z}_k^j \right)		
		+\Wk^n \, \vekt{Z}_k^n  \underbrace{ \prod_{j=n+1}^{n}  \left( \vekt{I} - \vekt{V}_k^j \vekt{Z}_k^j \right) }_{ =1 } \\
		 \nonumber 
&=   \JacN{0}  \prod_{i=0}^{n} \left( \vekt{I} - \vekt{V}_k^i \vekt{Z}_k^i \right) + \sum_{i=0}^{n} \vekt{W}_k^i \vekt{Z}_k^i  \prod_{j=i+1}^{n} \left( \vekt{I} - \vekt{V}_k^j \vekt{Z}_k^j \right)	~.
	\qquad \square
	\end{align}
\end{enumerate}

 \section{Implicit Jacobian Multiplication} \label{Appendix:JacobianProduct}
 
Based on the new Jacobian update formulation introduced in Section \ref{SubSec:IQN-IMVLS} and proven in \ref{AppendixProofA},
the product of the previous Jacobian approximation with a (residual) vector $\vekt{R}^k$ can be
evaluated as
\begin{align}
 	\JacN{n} \vekt{R}^k \approx  \sum_{i=n-q}^{n-1} \vekt{W}_k^i \vekt{Z}_k^i  \prod_{j=i+1}^{n-1}  \left( \vekt{R}^k - \vekt{V}_k^j \vekt{Z}_k^j  \vekt{R}^k \right) ~,
\end{align}
where $q$ of the $n$ past time steps are considered.
Note that the initial choice $\JacN{0}=\vekt{0}$ has been exploited.
The key for an efficient evaluation of this expression is to realize that the term
$\left( \vekt{I} - \vekt{V}_k^j \vekt{Z}_k^j \right)$
is needed for all $i<j$.
As a consequence, 
when looping over the previous time steps in reversed order,
it is sufficient to update the product term  $\prod_{j=i+1}^n \left( \vekt{R}^k - \vekt{V}_k^j \vekt{Z}_k^j  \vekt{R}^k \right)$ in every iteration $i$ by one product --
rather than evaluating it again and again.
This way, the overall computational complexity of the Jacobian-vector product 
evaluation reduces to $\mathcal{O}(m \bar{k} q)$,
where $\bar{k}$ is the average number of coupling iterations per time step.
For more details on the procedure and the cost of each step involved,
a pseudo-code realization is presented in Algorithm \ref{Alg:COmplexity}.
\begin{algorithm}[h!] 
	 \hrulefill \raisebox{-.8ex}{ Implicit Jacobian-Vector Product $\vekt{b}^k = \JacN{n} \, \vekt{R}^k $ }  \hrulefill \\

		\begin{minipage}[t]{0.7\textwidth}
					\comment{\textbf{Initialization:}} \\
			\begin{tabularx} {\textwidth}{ l X}
			\comment{Auxiliary vector for $\Pi$-terms: \hspace{2cm} }	 	&$ \vekt{a}_\pi = \vekt{R}^k$ \\
			\comment{Result vector:} 	& $\vekt{b}^k  = \vekt{0}$ \\
		\end{tabularx}
		\comment{\textbf{Loop over last $q$ time steps:}}\For{$i=n-1,\cdots,n-q$}	 
			{  
				\begin{tabularx} {\linewidth}{ l X X}
					\comment{Add contribution of time step $i$ to result $\vekt{b}^k$:}  & $\vekt{b}^k = \vekt{b}^k - \Wk^i \vekt{Z}_k^i \,\vekt{a}_\pi$ \\
						\comment{Update $\vekt{a}_\pi$ with $\Pi$-term of time step $i$:}	& $\vekt{a}_\pi=  \vekt{a}_\pi - \Vk^i \vekt{Z}_k^i \,\vekt{a}_\pi$ \\
				\end{tabularx}
			} 				
		\end{minipage}
	~
		\begin{minipage}[t]{0.2\textwidth}
			\vspace{0.05cm}
			\textit{$m \to \mathcal{O}(m)$}\\
			\textit{$m \to \mathcal{O}(m)$}
			\vspace{0.03cm} \\
			\textit{$q$ iterations} \vspace{0.03cm}\\ 
			\textit{$2m(\bar{k}+2) \to \mathcal{O}(m \bar{k})$}\\
			\textit{$2m(\bar{k}+2) \to \mathcal{O}(m \bar{k})$}\\
		\end{minipage} \\
			\Vhrulefill \\
			 \text{Total complexity: $4m(\bar{k}+2)q \to \mathcal{O}(m \bar{k} q)$} \\
		\Vhrulefill \\
		
	\caption[Implicit Jacobian-Vector Product]{\footnotesize{Pseudo-code for computing the product of the previous time step's inverse Jacobian with a vector  in an implicit manner. The number of floating point operations and the resulting computational complexity are indicated in italic.}} \label{Alg:COmplexity}
\end{algorithm} 
\section*{Acknowledgement}
We gratefully acknowledge the computing time granted by the J\"ulich-Aachen Research Alliance JARA-HPC. 

\bibliographystyle{model1-num-names}
\bibliography{references.bib}

\begin{thebibliography}{46}
\expandafter\ifx\csname natexlab\endcsname\relax\def\natexlab#1{#1}\fi
\providecommand{\bibinfo}[2]{#2}
\ifx\xfnm\relax \def\xfnm[#1]{\unskip,\space#1}\fi
\bibitem[{Bogaers et~al.(2014)Bogaers, Kok, Reddy, and Franz}]{Bogaers_IQNMVJ}
\bibinfo{author}{A.~E. Bogaers}, \bibinfo{author}{S.~Kok},
  \bibinfo{author}{B.~D. Reddy}, \bibinfo{author}{T.~Franz},
\newblock \bibinfo{title}{{Quasi-Newton Methods for Implicit Black-Box FSI
  Coupling}},
\newblock \bibinfo{journal}{Computer Methods in Applied Mechanics and
  Engineering} \bibinfo{volume}{279} (\bibinfo{year}{2014})
  \bibinfo{pages}{113--132}.
\bibitem[{F\"orster(2007)}]{FoersterPhD}
\bibinfo{author}{C.~F\"orster}, \bibinfo{title}{{Robust Methods for
  Fluid-Structure Interaction with Stabilized Finite Elements}}, Ph.D. thesis,
  University of Stuttgart, \bibinfo{year}{2007}.
\bibitem[{F\"orster et~al.(2007)F\"orster, Wall, and Ramm}]{Wall_AddedMass}
\bibinfo{author}{C.~F\"orster}, \bibinfo{author}{W.~A. Wall},
  \bibinfo{author}{E.~Ramm},
\newblock \bibinfo{title}{{Artificial Added Mass Instabilities in Sequential
  Staggered Coupling of Nonlinear Structures and Incompressible Viscous
  Flows}},
\newblock \bibinfo{journal}{Computer Methods in Applied Mechanics and
  Engineering} \bibinfo{volume}{196} (\bibinfo{year}{2007})
  \bibinfo{pages}{1278--1293}.
\bibitem[{K{\"u}ttler and Wall(2008)}]{FSI_DynamicRelax}
\bibinfo{author}{U.~K{\"u}ttler}, \bibinfo{author}{W.~A. Wall},
\newblock \bibinfo{title}{{Fixed-Point Fluid--Structure Interaction Solvers
  with Dynamic Relaxation}},
\newblock \bibinfo{journal}{{Computational Mechanics, Springer}}
  \bibinfo{volume}{43} (\bibinfo{year}{2008}) \bibinfo{pages}{61--72}.
\bibitem[{Irons and Tuck(1969)}]{irons1969version}
\bibinfo{author}{B.~M. Irons}, \bibinfo{author}{R.~C. Tuck},
\newblock \bibinfo{title}{{A Version of the Aitken Accelerator for Computer
  Iteration}},
\newblock \bibinfo{journal}{International Journal for Numerical Methods in
  Engineering} \bibinfo{volume}{1} (\bibinfo{year}{1969})
  \bibinfo{pages}{275--277}.
\bibitem[{Gatzhammer(2014)}]{GatzhammerPhD}
\bibinfo{author}{B.~Gatzhammer}, \bibinfo{title}{{Efficient and Flexible
  Partitioned Simulation of Fluid-Structure Interactions}}, Ph.D. thesis,
  Technical University of Munich, \bibinfo{year}{2014}.
\bibitem[{Degroote et~al.(2010{\natexlab{a}})Degroote, Haelterman, Annerel,
  Bruggeman, and Vierendeels}]{degroote2010performance}
\bibinfo{author}{J.~Degroote}, \bibinfo{author}{R.~Haelterman},
  \bibinfo{author}{S.~Annerel}, \bibinfo{author}{P.~Bruggeman},
  \bibinfo{author}{J.~Vierendeels},
\newblock \bibinfo{title}{{Performance of Partitioned Procedures in
  Fluid--Structure Interaction}},
\newblock \bibinfo{journal}{Computers \& Structures} \bibinfo{volume}{88}
  (\bibinfo{year}{2010}{\natexlab{a}}) \bibinfo{pages}{446--457}.
\bibitem[{Degroote et~al.(2010{\natexlab{b}})Degroote, Souto-Iglesias,
  Van~Paepegem, Annerel, Bruggeman, and Vierendeels}]{degroote2010partitioned}
\bibinfo{author}{J.~Degroote}, \bibinfo{author}{A.~Souto-Iglesias},
  \bibinfo{author}{W.~Van~Paepegem}, \bibinfo{author}{S.~Annerel},
  \bibinfo{author}{P.~Bruggeman}, \bibinfo{author}{J.~Vierendeels},
\newblock \bibinfo{title}{{Partitioned Simulation of the Interaction between an
  Elastic Structure and Free Surface Flow}},
\newblock \bibinfo{journal}{Computer Methods in Applied Mechanics and
  Engineering} \bibinfo{volume}{199} (\bibinfo{year}{2010}{\natexlab{b}})
  \bibinfo{pages}{2085--2098}.
\bibitem[{Gerbeau and Vidrascu(2003)}]{gerbeau2003quasi}
\bibinfo{author}{J.-F. Gerbeau}, \bibinfo{author}{M.~Vidrascu},
\newblock \bibinfo{title}{{A Quasi-Newton Algorithm based on a Reduced Model
  for Fluid-Structure Interaction Problems in Blood Flows}},
\newblock \bibinfo{journal}{ESAIM: Mathematical Modelling and Numerical
  Analysis} \bibinfo{volume}{37} (\bibinfo{year}{2003})
  \bibinfo{pages}{631--647}.
\bibitem[{Van~Brummelen et~al.(2005)Van~Brummelen, Michler, and
  De~Borst}]{van2005interface}
\bibinfo{author}{E.~Van~Brummelen}, \bibinfo{author}{C.~Michler},
  \bibinfo{author}{R.~De~Borst},
\newblock \bibinfo{title}{{Interface-GMRES (R) Acceleration of Subiteration for
  Fluid-Structure-Interaction Problems}},
\newblock \bibinfo{journal}{Report DACS-05-001}  (\bibinfo{year}{2005}).
\bibitem[{Degroote et~al.(2009)Degroote, Bathe, and
  Vierendeels}]{PerformancePartitionedMonolithic}
\bibinfo{author}{J.~Degroote}, \bibinfo{author}{K.-J. Bathe},
  \bibinfo{author}{J.~Vierendeels},
\newblock \bibinfo{title}{{Performance of a New Partitioned Procedure Versus a
  Monolithic Procedure in Fluid-Structure Interaction}},
\newblock \bibinfo{journal}{Computers \& Structures} \bibinfo{volume}{87}
  (\bibinfo{year}{2009}) \bibinfo{pages}{793--801}.
\bibitem[{Scheufele and Mehl(2017)}]{scheufele2017robust}
\bibinfo{author}{K.~Scheufele}, \bibinfo{author}{M.~Mehl},
\newblock \bibinfo{title}{{Robust Multisecant Quasi-Newton Variants for
  Parallel Fluid-Structure Simulations---and Other Multiphysics Applications}},
\newblock \bibinfo{journal}{SIAM Journal on Scientific Computing}
  \bibinfo{volume}{39} (\bibinfo{year}{2017}) \bibinfo{pages}{404--433}.
\bibitem[{Lindner et~al.(2015)Lindner, Mehl, Scheufele, and
  Uekermann}]{ComparisonQuasiNewton}
\bibinfo{author}{F.~Lindner}, \bibinfo{author}{M.~Mehl},
  \bibinfo{author}{K.~Scheufele}, \bibinfo{author}{B.~Uekermann},
\newblock \bibinfo{title}{{A Comparison of Various Quasi-Newton Schemes for
  Partitioned Fluid-Structure Interaction}},
\newblock \bibinfo{journal}{Proceedings of 6th International Conference on
  Computational Methods for Coupled Problems in Science and Engineering,
  Venice}  (\bibinfo{year}{2015}) \bibinfo{pages}{1--12}.
\bibitem[{Degroote and Vierendeels(2011)}]{Degroote_MultiSolver}
\bibinfo{author}{J.~Degroote}, \bibinfo{author}{J.~Vierendeels},
\newblock \bibinfo{title}{{Multi-Solver Algorithms for the Partitioned
  Simulation of Fluid-Structure Interaction}},
\newblock \bibinfo{journal}{Computer Methods in Applied Mechanics and
  Engineering} \bibinfo{volume}{200} (\bibinfo{year}{2011})
  \bibinfo{pages}{2195--2210}.
\bibitem[{Haelterman et~al.(2016)Haelterman, Bogaers, Scheufele, Uekermann, and
  Mehl}]{haelterman2016improving}
\bibinfo{author}{R.~Haelterman}, \bibinfo{author}{A.~E. Bogaers},
  \bibinfo{author}{K.~Scheufele}, \bibinfo{author}{B.~Uekermann},
  \bibinfo{author}{M.~Mehl},
\newblock \bibinfo{title}{{Improving the Performance of the Partitioned QN-ILS
  Procedure for Fluid--Structure Interaction Problems: Filtering}},
\newblock \bibinfo{journal}{Computers \& Structures} \bibinfo{volume}{171}
  (\bibinfo{year}{2016}) \bibinfo{pages}{9--17}.
\bibitem[{Yibin and Ogden(2001)}]{ogden2001nonlinear}
\bibinfo{author}{F.~Yibin}, \bibinfo{author}{R.~Ogden},
  \bibinfo{title}{{Nonlinear Elasticity: Theory and Applications}}, volume
  \bibinfo{volume}{283}, \bibinfo{year}{2001}.
\bibitem[{Bathe(1996)}]{BatheFEM}
\bibinfo{author}{K.-J. Bathe}, \bibinfo{title}{{Finite Element Procedures}},
  \bibinfo{publisher}{TBS}, \bibinfo{year}{1996}.
\bibitem[{Hosters(2018)}]{NorbertPhD}
\bibinfo{author}{N.~Hosters}, \bibinfo{title}{{Spline-Based Methods for
  Fluid-Structure Interaction}}, Ph.D. thesis, RWTH Aachen University,
  \bibinfo{year}{2018}.
\bibitem[{Braun(2007)}]{CarstenBraun_2007}
\bibinfo{author}{C.~Braun}, \bibinfo{title}{{E}in modulares {V}erfahren f\"ur
  die numerische aeroelastische {A}nalyse von {L}uftfahrzeugen}, Ph.D. thesis,
  RWTH Aachen University, \bibinfo{year}{2007}.
\bibitem[{K\"uttler et~al.(2006)K\"uttler, F\"orster, and
  Wall}]{Kuttler_Partitioned}
\bibinfo{author}{U.~K\"uttler}, \bibinfo{author}{C.~F\"orster},
  \bibinfo{author}{W.~A. Wall},
\newblock \bibinfo{title}{{A Solution for the Incompressibility Dilemma in
  Partitioned Fluid-Structure Interaction with Pure Dirichlet Fluid Domains}},
\newblock \bibinfo{journal}{Computational Mechanics, Springer}
  \bibinfo{volume}{38} (\bibinfo{year}{2006}) \bibinfo{pages}{417--429}.
\bibitem[{Pauli(2016)}]{LutzPhD}
\bibinfo{author}{L.~Pauli}, \bibinfo{title}{{Stabilized Finite Element Methods
  for Computational Design of Blood-Handling Devices}}, Ph.D. thesis, RWTH
  Aachen University, \bibinfo{year}{2016}.
\bibitem[{Donea and Huerta(2003)}]{FEM_Flows}
\bibinfo{author}{J.~Donea}, \bibinfo{author}{A.~Huerta},
  \bibinfo{title}{{Finite Element Methods for Flow Problems}},
  \bibinfo{publisher}{WILEY}, \bibinfo{year}{2003}.
\bibitem[{Tezduyar and Behr(1992{\natexlab{a}})}]{ArticleDSDSST_1}
\bibinfo{author}{T.~E. Tezduyar}, \bibinfo{author}{M.~Behr},
\newblock \bibinfo{title}{{A New Strategy for Finite Element Computations
  Involving Moving Boundaries and Interfaces - The
  Deforming-Spatial-Domain/Space-Time Procedure: I. The Concept and the
  Preliminary Numerical Tests}},
\newblock \bibinfo{journal}{Computer Methods in Applied Mechanics and
  Engineering} \bibinfo{volume}{94} (\bibinfo{year}{1992}{\natexlab{a}})
  \bibinfo{pages}{353--371}.
\bibitem[{Tezduyar and Behr(1992{\natexlab{b}})}]{ArticleDSDSST_2}
\bibinfo{author}{T.~E. Tezduyar}, \bibinfo{author}{M.~Behr},
\newblock \bibinfo{title}{{A New Strategy for Finite Element Computations
  Involving Moving Boundaries and Interfaces - The
  Deforming-Spatial-Domain/Space-Time Procedure: II. Computation of
  Free-Surface Flows, Two-Liquid Flows, and Flows with Drifting Cylinders}},
\newblock \bibinfo{journal}{Computer Methods in Applied Mechanics and
  Engineering} \bibinfo{volume}{94} (\bibinfo{year}{1992}{\natexlab{b}})
  \bibinfo{pages}{353--371}.
\bibitem[{Elgeti and Sauerland(2016)}]{Steffi2015}
\bibinfo{author}{S.~Elgeti}, \bibinfo{author}{H.~Sauerland},
\newblock \bibinfo{title}{{Deforming Fluid Domains within the Finite Element
  Method: Five Mesh-Based Tracking Methods in Comparison}},
\newblock \bibinfo{journal}{Archives of Computational Methods in Engineering}
  \bibinfo{volume}{23} (\bibinfo{year}{2016}) \bibinfo{pages}{323--361}.
\bibitem[{Behr and Abraham(2002)}]{Behr_Abraham}
\bibinfo{author}{M.~Behr}, \bibinfo{author}{F.~Abraham},
\newblock \bibinfo{title}{{F}ree-{S}urface {F}low {S}imulation in the
  {P}resence of {I}nclined {W}alls},
\newblock \bibinfo{journal}{Computer Methods in Applied Mechanics and
  Engineering} \bibinfo{volume}{191} (\bibinfo{year}{2002})
  \bibinfo{pages}{5467--5483}.
\bibitem[{Cottrell et~al.(2009)Cottrell, Hughes, and Bazilevs}]{IGA_Hughes}
\bibinfo{author}{J.~A. Cottrell}, \bibinfo{author}{T.~J.~R. Hughes},
  \bibinfo{author}{Y.~Bazilevs}, \bibinfo{title}{{Isogeometric Analysis -
  Toward Integration of CAD and FEA}}, \bibinfo{publisher}{WILEY},
  \bibinfo{year}{2009}.
\bibitem[{Chung and Hulbert(1993)}]{Bossak_Chung}
\bibinfo{author}{J.~Chung}, \bibinfo{author}{G.~M. Hulbert},
\newblock \bibinfo{title}{{A Time Integration Algorithm for Structural Dynamics
  with Improved Numerical Dissipation: The Generalized-\ensuremath{\alpha}
  Method}},
\newblock \bibinfo{journal}{Journal of Applied Mechanics} \bibinfo{volume}{60}
  (\bibinfo{year}{1993}) \bibinfo{pages}{371--375}.
\bibitem[{Kuhl and Crisfield(1999)}]{Bossak_Kuhl}
\bibinfo{author}{D.~Kuhl}, \bibinfo{author}{A.~Crisfield},
\newblock \bibinfo{title}{{Energy-Conserving and Decaying Algorithms in
  Non-Linear Structural Dynamics}},
\newblock \bibinfo{journal}{International Journal for Numerical Methods in
  Engineering} \bibinfo{volume}{45} (\bibinfo{year}{1999})
  \bibinfo{pages}{569--599}.
\bibitem[{Hughes et~al.(2005)Hughes, Cottrell, and Bazilevs}]{IGA_2004}
\bibinfo{author}{T.~J.~R. Hughes}, \bibinfo{author}{J.~A. Cottrell},
  \bibinfo{author}{Y.~Bazilevs},
\newblock \bibinfo{title}{{I}sogeometric {A}nalysis: {CAD}, {F}inite
  {E}lements, {NURBS}, {E}xact {G}eometry and {M}esh {R}efinement},
\newblock \bibinfo{journal}{Computer Methods in Applied Mechanics and
  Engineering} \bibinfo{volume}{194} (\bibinfo{year}{2005})
  \bibinfo{pages}{4135--4195}.
\bibitem[{Piegl and Tiller(2012)}]{thenurbsbook}
\bibinfo{author}{L.~Piegl}, \bibinfo{author}{W.~Tiller}, \bibinfo{title}{{The
  NURBS Book}}, \bibinfo{publisher}{Springer Science \& Business Media},
  \bibinfo{year}{2012}.
\bibitem[{Ramm and Wall(2004)}]{Shells_SensitiveRelation}
\bibinfo{author}{E.~Ramm}, \bibinfo{author}{W.~A. Wall},
\newblock \bibinfo{title}{Shell {S}tructures - {A} {S}ensitive {I}nterrelation
  between {P}hysics and {N}umerics},
\newblock \bibinfo{journal}{International Journal for Numerical Methods in
  Engineering} \bibinfo{volume}{60} (\bibinfo{year}{2004})
  \bibinfo{pages}{381--427}.
\bibitem[{Dornisch(2015)}]{Dornisch_PhD}
\bibinfo{author}{W.~Dornisch}, \bibinfo{title}{{I}nterpolation of {R}otations
  and {C}oupling of {P}atches in {I}sogeometric {R}eissner-{M}indlin {S}hell
  {A}nalysis}, Ph.D. thesis, RWTH Aachen University, \bibinfo{year}{2015}.
\bibitem[{Dornisch et~al.(2013)Dornisch, Klinkel, and Simeon}]{Dornisch_A}
\bibinfo{author}{W.~Dornisch}, \bibinfo{author}{S.~Klinkel},
  \bibinfo{author}{B.~Simeon},
\newblock \bibinfo{title}{{Isogeometric Reissner-Mindlin Shell Analysis with
  Exactly Calculated Director Vectors}},
\newblock \bibinfo{journal}{Computer Methods in Applied Mechanics and
  Engineering} \bibinfo{volume}{253} (\bibinfo{year}{2013})
  \bibinfo{pages}{491--504}.
\bibitem[{Dornisch and Klinkel(2014)}]{Dornisch_D}
\bibinfo{author}{W.~Dornisch}, \bibinfo{author}{S.~Klinkel},
\newblock \bibinfo{title}{{Treatment of Reissner-Mindlin Shells with Kinks
  without the Need for Drilling Rotation Stabilization in an Isogeometric
  Framework}},
\newblock \bibinfo{journal}{Computer Methods in Applied Mechanics and
  Engineering} \bibinfo{volume}{276} (\bibinfo{year}{2014})
  \bibinfo{pages}{35--66}.
\bibitem[{Hosters et~al.(2018)Hosters, Helmig, Stavrev, Behr, and
  Elgeti}]{Hosters2017}
\bibinfo{author}{N.~Hosters}, \bibinfo{author}{J.~Helmig},
  \bibinfo{author}{A.~Stavrev}, \bibinfo{author}{M.~Behr},
  \bibinfo{author}{S.~Elgeti},
\newblock \bibinfo{title}{{Fluid--Structure Interaction with NURBS-based
  Coupling}},
\newblock \bibinfo{journal}{Computer Methods in Applied Mechanics and
  Engineering} \bibinfo{volume}{332} (\bibinfo{year}{2018})
  \bibinfo{pages}{520--539}.
\bibitem[{Causin et~al.(2005)Causin, Gerbeau, and Nobile}]{CausinAddedMass}
\bibinfo{author}{P.~Causin}, \bibinfo{author}{J.-F. Gerbeau},
  \bibinfo{author}{F.~Nobile},
\newblock \bibinfo{title}{{Added-Mass Effect in the Design of Partitioned
  Algorithms for Fluid-Structure Problems}},
\newblock \bibinfo{journal}{Computer Methods in Applied Mechanics and
  Engineering} \bibinfo{volume}{194} (\bibinfo{year}{2005})
  \bibinfo{pages}{4506--4527}.
\bibitem[{Van~Brummelen(2009)}]{BrummelenAddedMass}
\bibinfo{author}{E.~H. Van~Brummelen},
\newblock \bibinfo{title}{{Added Mass Effects of Compressible and
  Incompressible Flows in Fluid-Structure Interaction}},
\newblock \bibinfo{journal}{Journal of Applied Mechanics} \bibinfo{volume}{76}
  (\bibinfo{year}{2009}) \bibinfo{pages}{021206}.
\bibitem[{Uekermann(2016)}]{uekermann2016partitioned}
\bibinfo{author}{B.~W. Uekermann}, \bibinfo{title}{{Partitioned Fluid-Structure
  Interaction on Massively Parallel Systems}}, Ph.D. thesis, Technical
  University of Munich, \bibinfo{year}{2016}.
\bibitem[{Bogaers et~al.(2012)Bogaers, Kok, and Franz}]{IQN_POD}
\bibinfo{author}{A.~E. Bogaers}, \bibinfo{author}{S.~Kok},
  \bibinfo{author}{T.~Franz},
\newblock \bibinfo{title}{{Strongly Coupled Partitioned FSI Using Proper
  Orthogonal Decomposition}},
\newblock \bibinfo{journal}{Conference Paper, Eighth South African Conference
  on Computational and Applied Mechanics (SACAM)}  (\bibinfo{year}{2012}).
\bibitem[{Golub and Van~Loan(2012)}]{golub1996matrix}
\bibinfo{author}{G.~H. Golub}, \bibinfo{author}{C.~F. Van~Loan},
  \bibinfo{title}{{Matrix Computations}}, \bibinfo{publisher}{JHU Press},
  \bibinfo{year}{2012}.
\bibitem[{Scheufele(2018)}]{ScheufelePhD}
\bibinfo{author}{K.~Scheufele}, \bibinfo{title}{{Coupling Schemes and Inexact
  Newton for Multi-Physics and Coupled Optimization Problems}}, Ph.D. thesis,
  University of Stuttgart, \bibinfo{year}{2018}.
\bibitem[{Farmaga et~al.(2011)Farmaga, Shmigelskyi, Spiewak, and
  Ciupinski}]{farmaga2011evaluation}
\bibinfo{author}{I.~Farmaga}, \bibinfo{author}{P.~Shmigelskyi},
  \bibinfo{author}{P.~Spiewak}, \bibinfo{author}{L.~Ciupinski},
\newblock \bibinfo{title}{{Evaluation of Computational Complexity of Finite
  Element Analysis}},
\newblock \bibinfo{journal}{11th International Conference The Experience of
  Designing and Application of CAD Systems in Microelectronics (CADSM)}
  (\bibinfo{year}{2011}) \bibinfo{pages}{213--214}.
\bibitem[{Graham and Adler(2006)}]{graham2006nodal}
\bibinfo{author}{B.~Graham}, \bibinfo{author}{A.~Adler},
\newblock \bibinfo{title}{{A Nodal Jacobian Inverse Solver for Reduced
  Complexity EIT Reconstructions}},
\newblock \bibinfo{journal}{International Journal of Information and Systems
  Sciences} \bibinfo{volume}{2} (\bibinfo{year}{2006})
  \bibinfo{pages}{453--468}.
\bibitem[{Zhou and Jiao(2013)}]{zhou2013linear}
\bibinfo{author}{B.~Zhou}, \bibinfo{author}{D.~Jiao},
\newblock \bibinfo{title}{{A Linear Complexity Direct Finite Element Solver for
  Large-scale 3-D Electromagnetic Analysis}},
\newblock \bibinfo{journal}{2013 IEEE Antennas and Propagation Society
  International Symposium (APSURSI)}  (\bibinfo{year}{2013})
  \bibinfo{pages}{1684--1685}.
\bibitem[{Greisen et~al.(2013)Greisen, Runo, Guillet, Heinzle, Smolic, Kaeslin,
  and Gross}]{greisen2013evaluation}
\bibinfo{author}{P.~Greisen}, \bibinfo{author}{M.~Runo},
  \bibinfo{author}{P.~Guillet}, \bibinfo{author}{S.~Heinzle},
  \bibinfo{author}{A.~Smolic}, \bibinfo{author}{H.~Kaeslin},
  \bibinfo{author}{M.~Gross},
\newblock \bibinfo{title}{{Evaluation and FPGA Implementation of Sparse Linear
  Solvers for Video Processing Applications}},
\newblock \bibinfo{journal}{IEEE Transactions on Circuits and Systems for Video
  Technology} \bibinfo{volume}{23} (\bibinfo{year}{2013})
  \bibinfo{pages}{1402--1407}.

\end{thebibliography}

\end{document}